\documentclass{aa}  

\usepackage{graphicx}
\usepackage{xcolor}
\usepackage{txfonts}
\usepackage{lipsum}
\usepackage{hyperref}
\usepackage{subcaption}         
\usepackage{lscape}             
\usepackage{placeins}           
                                
\begin{document}

   \title{Contributions of massive stars and classical novae to Galactic \nuc{Al}{26}}


%

    \author{Manja\,L. Zimmerer\inst{1}
        \and Thomas Siegert\inst{1}
        \and Niklas\,C. Bauer\inst{1}
        \and Tristan Bouchet\inst{1}
        \and Patrik Ehrmann\inst{1}
        \and Mika\,A. Gelowicz\inst{1,}\inst{2}
        \and Laura Eisenberger\inst{1}
        \and Saurabh Mittal\inst{1}
        \and Tomohiko Oka\inst{1}
        \and Rudi Reinhardt\inst{1}
        \and Manuel\,R.\,H.\,W. Skalka\inst{1}
        \and Dimitris Tsatsis \inst{1}
        \and Hiroki Yoneda \inst{3,4,5,6}
        \and Arianna Vasini \inst{7}
        }

   \institute{Julius-Maximilians-Universität Würzburg, Fakultät für Physik und Astronomie, Institut für Theoretische Physik und Astrophysik, Lehrstuhl für Astronomie, Emil-Fischer-Str. 31, D-97074 Würzburg, Germany
   \and
   Rheinische Friedrich-Wilhelms-Universität Bonn, Argelander-Institut für Astronomie (AIfA), Auf dem Hügel 71, D-53121 Bonn, Germany
   \and
   The Hakubi Center for Advanced Research, Kyoto University, Yoshida Ushinomiyacho, Sakyo-ku, Kyoto 606-8501, Japan
   \and
   Department of Physics, Kyoto University, Kitashirakawa Oiwake-cho, Sakyo-ku, Kyoto 606-8502, Japan
   \and
   RIKEN Nishina Center, 2-1 Hirosawa, Wako, Saitama 351-0198, Japan
   \and
   Kavli Institute for the Physics and Mathematics of the Universe (WPI), UTIAS, The University of Tokyo, 5-1-5 Kashiwanoha, Kashiwa, Chiba 277-8583, Japan,
   \and
   Como Lake centre for AstroPhysics (CLAP), DiSAT, Universit{\`a} dell’Insubria, via Valleggio 11, 22100 Como, Italy
   }

   \date{Received XX , 20XX}

 
  \abstract
   {The radioactive isotope \nuc{Al}{26} is believed to be mainly produced by massive stars and ejected into the interstellar medium through their winds and supernovae.
   Due to its lifetime of 1.04\,Myr, it traces recent nucleosynthesis and star formation.
   \nuc{Fe}{60} is a similarly long-lived (lifetime 3.8\,Myr) isotope, and only originating in ccSNe.
   Stellar evolution models in combination with Galactic Chemical Evolution (GCE) models tend to over-predict the $\nuc{Fe}{60}/\nuc{Al}{26}$ ratio.
   Population synthesis models that attempt to explain both, the \nuc{Al}{26} mass in the entire Milky Way and its spatial distribution, by comparisons to the 1.8\,MeV decay $\gamma$-ray line, only match the observational constraints with an enhanced star formation rate (SFR) of $\gtrsim 5\,\mathrm{M_\odot\,yr^{-1}}$. 
   Classical novae may contribute significantly to the \nuc{Al}{26} production, potentially lowering the SFR inferred from massive stars, 
   even though previous studies suggest their contribution is negligible or $\leq$\,10\%. However, a large nova contribution would also alleviate some tension in the \nuc{Fe}{60}/\nuc{Al}{26} ratio.}
   {In this work we aim to determine the mass distribution of \nuc{Al}{26} as a function of Galactocentric radius and estimate the contribution of (very) massive stars and classical novae to the 1.8\,MeV signal.} 
   {We are using 20\,yr of INTEGRAL/SPI observations of the $\gamma$-ray line at $1.8\,\mathrm{MeV}$ to test different assumptions on the radial structure of \nuc{Al}{26} and compare the resulting mass distribution to different GCE models.}
    {For our baseline analysis, all tested source models prefer substantial nova contributions, with best-fit fractions of 77--84\%.
   Because the nova and massive-star amplitudes are anticorrelated, the full two-parameter $2\sigma$ confidence regions permit substantially smaller nova fractions.
  
   If the preferred large nova contribution is adopted, this alleviates tension in the \nuc{Fe}{60}/\nuc{Al}{26} ratio and lowers the SFR from \nuc{Al}{26} $\gamma$-ray measurements accordingly to $1$--$2\,\mathrm{M_\odot\,yr^{-1}}$.}
   {Our baseline fits favour a substantial contribution from classical novae, but the quantitative lower bound depends strongly on the adopted source profile and on the treatment of the Solar neighbourhood. Under some tested assumptions, little or no nova contribution is permitted at the $2\sigma$ level. While it is difficult to gauge both contributions directly, next generation MeV telescopes can potentially identify individual massive stars, such as $\gamma^2$-Velorum.}

   \keywords{Stars: novae -- Stars: massive -- Gamma rays: ISM
               }

   \maketitle
    \nolinenumbers 

\section{Introduction}\label{introduction}
\begin{figure} [t!]
     \centering
     \includegraphics[width=\linewidth,trim=0.0cm 0.1cm 0.0cm 0.0cm,clip=True]{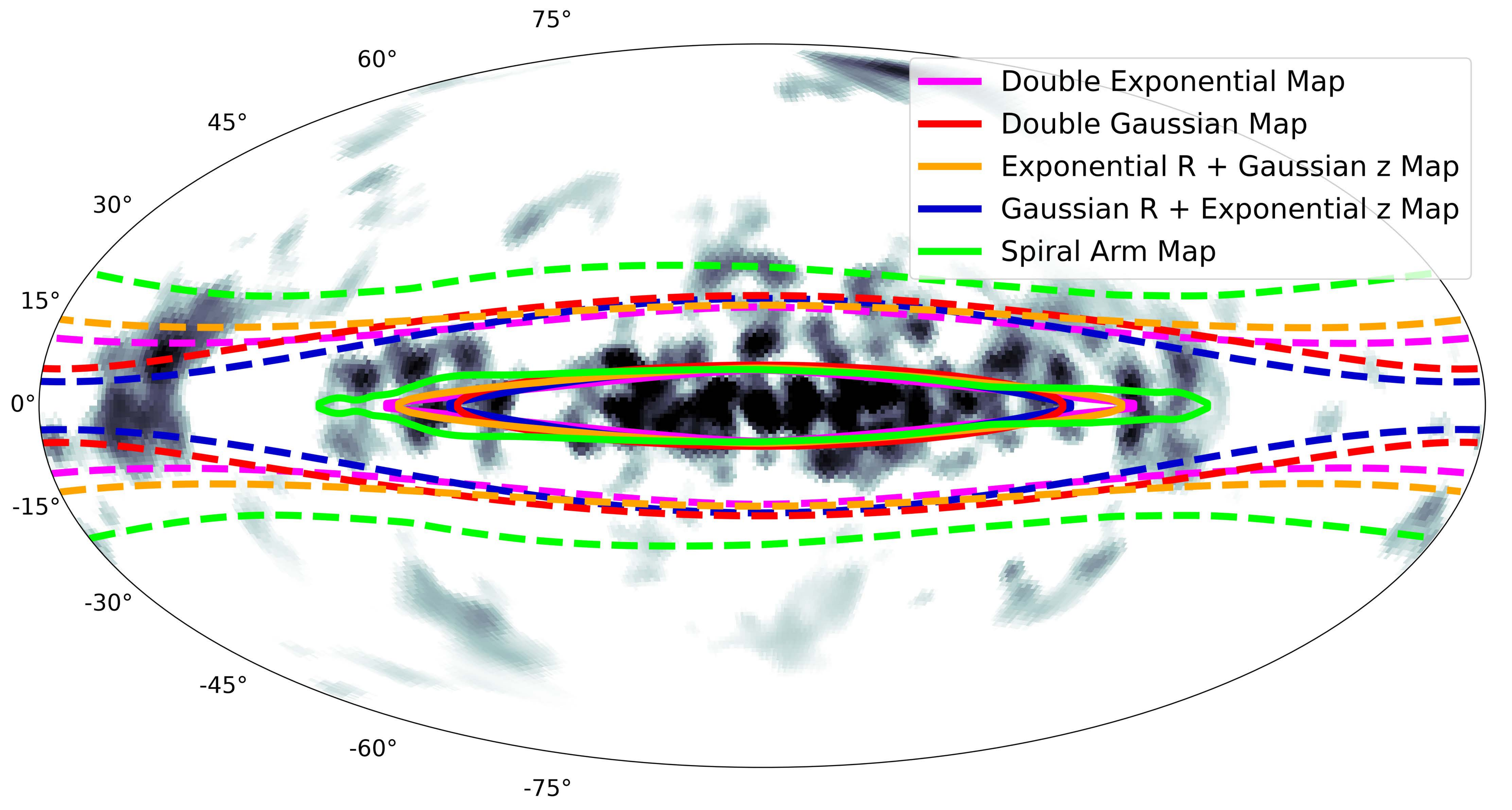}
     
     \caption{All sky map in Galactic coordinates of \nuc{Al}{26} emission imaged with INTEGRAL/SPI \citep{Bouchet2015_SPI} overlaid by contours of different Galactic disk models with scale height $h_E$=0.6\,kpc and scale radius $R_E$=5.0\,kpc as a best-fit empirical model. {The contours show levels of 80\%(solid line) and 95\%(dashed line) model emissivity.} }
     \label{SkyMap}
 \end{figure}

The radioactive isotope \nuc{Al}{26} is characterized by its  $1.809$\,MeV $\gamma$-ray line. Due to its intermediate lifetime of $\tau=1.04$\,Myr, it is considered to be an excellent tracer of the ongoing nucleosynthesis in our Galaxy \citep{Diehl2006_26Al}. It is long-lived enough to be ejected from stars into the interstellar medium (ISM), either during the wind phase or through core-collapse supernovae (ccSNe), before decaying. Compared to the timescale of Galactic evolution, the lifetime is still small, so that we only observe recently produced \nuc{Al}{26}.
The most recent 1.8\,MeV all-sky map from measurements with INTEGRAL/SPI \citep{Winkler_INTEGRAL_2003, Verdrenne_SPI_2003}, Fig.\,\ref{SkyMap} \citep{Bouchet2015_SPI}, shows extended emission along the Galactic plane, supporting the consensus idea that the isotope originates in regions of massive-star groups, based on the clumpy morphology that is coincident with nearby OB associations \citep{Knoedlsedler_1999}. 

The isotope \nuc{Fe}{60} ($\gamma$-ray lines at 1173 and 1332\,keV from the \nuc{Co}{60} decay, and at 59\,keV from the \nuc{Fe}{60} decay directly) has a similar lifetime of $3.8\,$Myr and is only expected to be ejected in ccSNe \citep[e.g.,][]{Heger2000_MS,Wang_2020}. Because of their similar origin and lifetimes, the ratio \nuc{Fe}{60}/\nuc{Al}{26} of their $\gamma$-ray emission is thought to be independent of the true location and distribution of their sources, and provides a chance to test stellar evolution models \citep{Timmes_1995}. However, the measured \nuc{Fe}{60}/\nuc{Al}{26} flux ratio of $0.182\pm0.042$ \citep{Wang_2020}, is usually overshot by a factor of 3--10 in model calculations \citep[e.g.,][]{Spyrou_nature, Sukhbold_2016, Austin_ratio_2017}. 
This method, however, is only valid if \nuc{Al}{26} and \nuc{Fe}{60} truly have the same origins throughout the Galaxy.

At the same time, \nuc{Al}{26} measurements can be used to estimate the star formation rate (SFR) in the Milky Way, resulting in  $\gtrsim 5\,\mathrm{M_\odot\,yr^{-1}}$ \citep{Siegert2023_GalacticPopulation, dissertation_Pleintinger}, while other measurements only suggest a SFR of $1$--$2\,\mathrm{M_\odot\,yr^{-1}}$ \citep[e.g.,][]{SFRother, Licquia_2015_SFR, Chomiuk2011_SFR}.

In both cases, tensions could be alleviated if other sources also produce considerable amounts of \nuc{Al}{26}. All environments of hydrogen burning are potential sources of \nuc{Al}{26}, as it is fused in the \(^{25}\text{Mg}(\text{p},\gamma)^{26}\text{Al}\) reaction \citep{Jose2006}. 
All main sequence stars experience hydrogen burning, but more massive stars above $10$--$25\,\mathrm{M_\odot}$ evolve quickly, on a timescale of $\sim 10\,\mathrm{Myr}$, through the main sequence and still contain a significant amount of \nuc{Al}{26} when entering later burning stages. This phase is then accompanied by mass loss with strong stellar winds. Stars in this mass range are typically considered to end their lifes via ccSNe (SN). 
However, the typically assumed ccSN for all massive stars above $8\,\mathrm{M_\odot}$ may, in fact, not be true. In \citet{Sukhbold_2016}, different model assumptions have been tested in 1D simulations. It was found that there could be ``islands'' of exploding stars as a function of mass, and intervals of initial masses that collapse directly into black holes without explosion. Other authors \citep[e.g.,][]{Limongi_Chieffi_18} assume that no star with an initial mass above $25\,\mathrm{M_\odot}$ actually explodes, which may be corroborated by observations \citep{Smartt_2009}.
Only stars that explode would significantly contribute to the \nuc{Al}{26} budget and enhance the persistent Galactic mass. Although wind yields for higher mass stars would remain unchanged, their ccSN yields would go to zero.

Another prime candidate for \nuc{Al}{26} are classical novae (CNe) \citep{Gomez-Gomar}. These surface explosions occur in binary systems, in which a white dwarf (WD) accretes hydrogen-rich matter from a companion star, usually a late-type main-sequence star \citep{Darnley_2012_novaeprogen}. The accreted matter closest to the surface gets compressed by the matter above and degenerates over time. {At that point, the pressure is only weakly dependent on the temperature, and the degenerate matter is not able to respond to heat  
by expansion anymore. This leads to fusion processes, including hydrogen burning, being ignited in a thermonuclear runaway reaction. Hydrogen burning in such a thin accreted shell is thermally unstable, since the nuclear energy generation rate is extremely sensitive to temperature increases. Because of the suppressed expansion, this amplifies the runaway reaction even further.} Heated by the fusion processes, at some point, the system is hot enough for the degeneracy line to be crossed, and the matter expands rapidly, ejecting the outer layer.  
This is especially interesting as the nova rate at \(R_{\text{CN}}={50}_{-23} ^{+31}\,\text{yr}^{-1}\) \citep{Shafter17_GalacticNovarate} in the Galaxy is much higher than the ccSNe rate, with only \(R_{\text{SN}}=(0.018-0.028)\,\text{yr}^{-1}\) \citep[e.g.,][]{Siegert2023_GalacticPopulation}. Only one third of these novae happen on oxygen-neon (ONe) WDs, containing the seed nuclei needed for substantial synthesis of \nuc{Al}{26} \citep{Jose2007}.
The expected persistent mass of \nuc{Al}{26} from novae and massive stars in the Galaxy can be expressed as the product of the respective rate, the average yield of the event $\langle Y_{26} \rangle$, and the lifetime of the isotope.
We therefore expect a mass ratio of
\begin{align}
    \frac{M_{\text{SN}}}{M_{\text{CN}}}=\frac{R_{\text{SN}}}{R_{\text{CN}}}\frac{\langle Y_{26}\rangle_{\text{SN}}}{\langle Y_{26}\rangle_{\text{CN}}}.
\end{align}
\citet{Starrfield_yieldsONe} present the outcome of hydrodynamic simulations of ONe novae nucleosynthesis and predict yields between $3.7\times10^{-11}$\,M$_\odot$ and $2.1\times10^{-7}$\,M$_\odot$. Yields of massive stars were presented by \citet{Limongi_Chieffi_18}, and are between $1\times10^{-5}$\,M$_\odot$ and $1.8\times10^{-3}$\,M$_\odot$ for stars of initial masses from 8 to 120\,M$_\odot$ \citep[see also][for very massive star yields; cf. Sect.\,\ref{sec2}]{Higgins2023_VMS}.
By taking into account the range of all possibilities, we expect $\log_{10}\left(M_{\mathrm{SN}} / M_{\mathrm{CN}} \right) =2.2\pm3.0$. 
However, if no star with initial mass of more than 25\,M$_\odot$ actually explodes, the typical yield of massive stars decreases by one order of magnitude, and the expected mass ratio is $\log_{10}( \frac{M_{\text{SN}}}{M_{\text{CN}}})=1.6\pm2.4$. This means that, given the uncertainties of different massive star yield and nova yield models, novae might even contribute more to the Galactic \nuc{Al}{26} abundance than massive stars.

With the sensitivity and angular resolution of INTEGRAL/SPI, it is difficult to measure individual yields from massive stars or single novae. However, there was one observation of a molecule including \nuc{Al}{26} in a remnant of a potential nova-like event. Radioactive aluminium monofluoride (\nuc{AlF}{26}) has been found in CK Vul (or Nova 1670) utilizing ALMA data \citep{Kaminski_26AlF}.

However, it remains difficult to determine the absolute mass of ejecta from atomic or molecular transitions.
Instead of relying on the yields of various calculations, we attempt to decipher the contributions of massive stars and novae by their expected {relative} geometrical distribution in the Milky Way. While it is almost certain that the yields from novae would distribute inside superbubbles, similar to those from massive stars, the contribution of novae as an old population of stars compared to massive stars as a much younger population would imprint in the radial distribution. We use this geometrical approach to determine absolute masses as a function of Galactocentric radius for four different geometrical models and compare them to {relative} expectations of \nuc{Al}{26} from novae and massive stars.
The absolute yield normalizations per source type are therefore largely irrelevant to this fitting approach because we fit the normalisation of each source population to match the observed flux.
The decomposition is instead limited by the accuracy of the adopted relative source-distribution profiles.

\begin{figure}[h!]
    \centering
    \includegraphics[width=\linewidth]{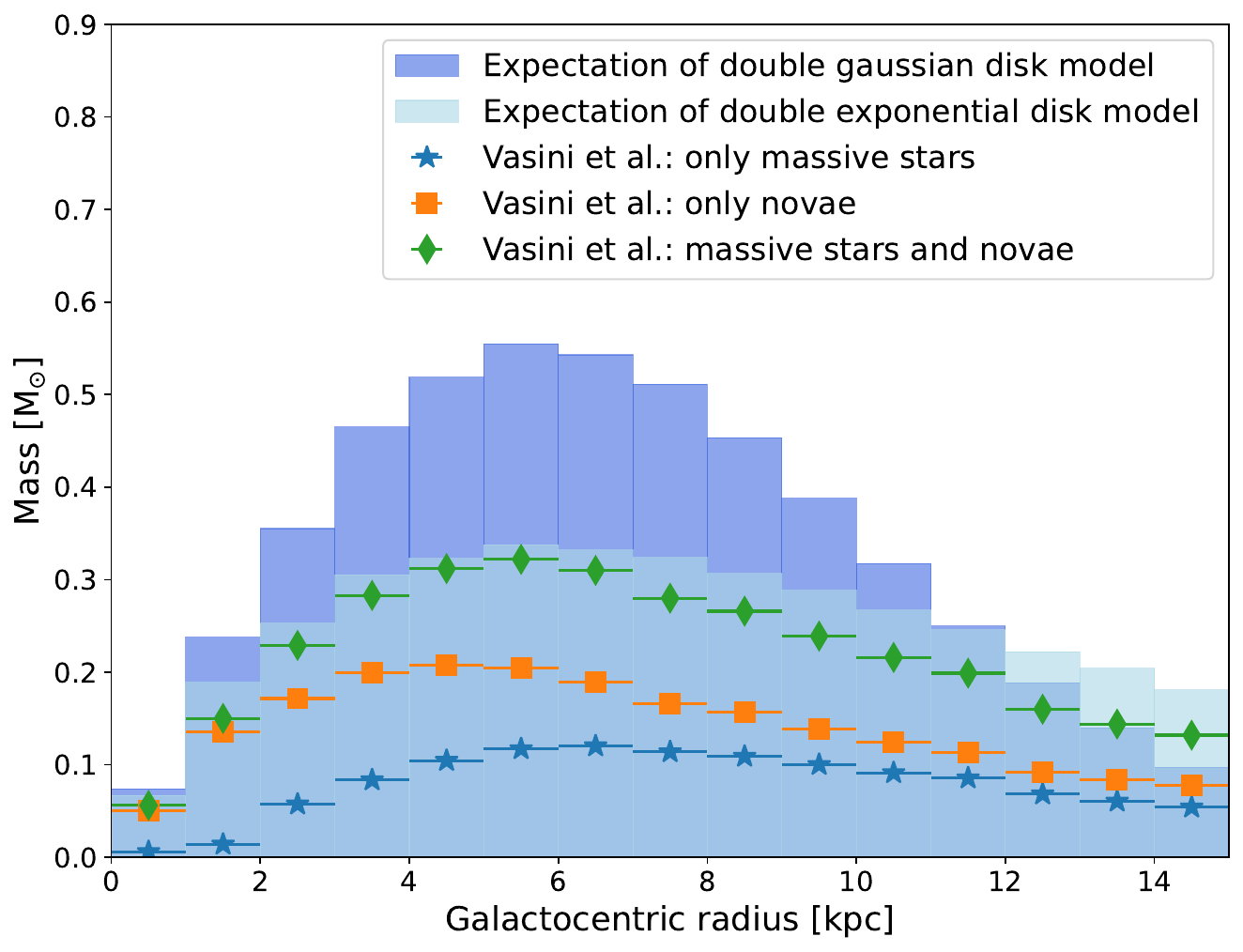}
    \caption{Comparison of the mass values predicted by \cite{vasini2024tracingmilkywayspiral}, and the geometric distributions of masses predicted by a double exponential model and a double gaussian model. While they peak at the same distance from the Galactic centre, their relative distribution is  different. The mass originating only from classical novae (orange) and only from massive stars (blue) sum to the total prediction (green). The blue histogram shows the masses inside rings with fixed scale height and radius up to 15\,kpc in 1\,kpc steps.} 
    \label{Vasinivsus}
\end{figure}

This paper is structured as follows:
In Sect.\,\ref{sec2}, we describe the theoretical predictions of the \nuc{Al}{26} distribution depending on the source population throughout the Galaxy for different Galactic Chemical Evolution (GCE) models that take the geometry of the Galaxy into account. Sect.\,\ref{INTEGRAL/SPI_analysis} contains a description of our dataset and the INTEGRAL/SPI analysis. In Sect.\,\ref{galactic_disk_models}, we show the spatial modelling of the Galactic disk at 1.8\,MeV and present the observed distribution of \nuc{Al}{26}. We estimate the contributions of massive stars and novae to Galactic \nuc{Al}{26} in Sect.\,\ref{contribution_results}, and close with a summary and conclusion in Sect.\,\ref{summary}.

\section{Predictions from Galactic chemical evolution models}\label{sec2}
\citet{vasini2024tracingmilkywayspiral} published their results of a two-dimensional GCE model of our Galaxy, with the aim to predict the \nuc{Al}{26} mass as a function of Galactocentric radius. As an input they used the SFR, considering a spiral perturbation to account for the spiral arms of the Milky Way, and the nova rate. 
For novae, they modeled the yields according to \citet{Jose2007} and considered that the nova population consists of 70\% CO novae and 30\% ONe novae. As the yield depends on the initial mass of the WD, \citet{vasini2024tracingmilkywayspiral} chose the average of those provided in \citet{Jose2007}, based on \citet{Romano2003}. The yields of massive stars were the ones presented in \citet{Kobayashi2006} combined with the results for \nuc{Al}{26} from \citet{Woosley195}. 
Even though the SFR of the disk peaked around \(10\,\)Gyr ago, the spiral pattern of the arms can still be distinguished in \nuc{Al}{26}. 

The first novae only happened after the disk SFR had already peaked, as the binary systems first needed to evolve. This is causing a delay between the nova rate and the SFR and is the reason why the SFR cannot be traced by \nuc{Al}{26} alone. They used their GCE model to test two different scenarios: while the first one (hereafter Model\,1) only considers the contribution of mostly massive stars, with minor contributions of type Ia supernovae (SNe Ia) and asymptotic giant branch (AGB) stars, Model\,2 additionally includes novae as a source of \nuc{Al}{26}. The distribution throughout the Milky Way provided by these models, up to a radius of 15\,kpc, differs by the total amount of \nuc{Al}{26} and its spatial distribution.

While for Model\,1, the distribution still follows the spiral arm pattern, in Model\,2, the spiral arms cannot be distinguished easily any more. The amount of  \nuc{Al}{26} produced by novae smears out the correlation between the SFR and \nuc{Al}{26}. 
In \cite{vasini2024tracingmilkywayspiral}, the total mass integrated up to 5\,kpc is given as 0.3\,M$_\odot$ for Model\,1 and 1.0\,M$_\odot$ for Model\,2, so that the total mass over this region is around four times as large when novae are considered as sources of \nuc{Al}{26}, or put differently: massive stars only contribute around 25\% of the \nuc{Al}{26} mass.
This statement holds even in the case of the full Galaxy model up to 15\,kpc, with masses of 2.1\,M$_\odot$ from novae, 1.2\,M$_\odot$ from massive stars, for a total mass of 3.3\,M$_\odot$ \nuc{Al}{26} in the Milky Way.
\begin{figure}[!ht]
    \centering
    \includegraphics[width=\linewidth]{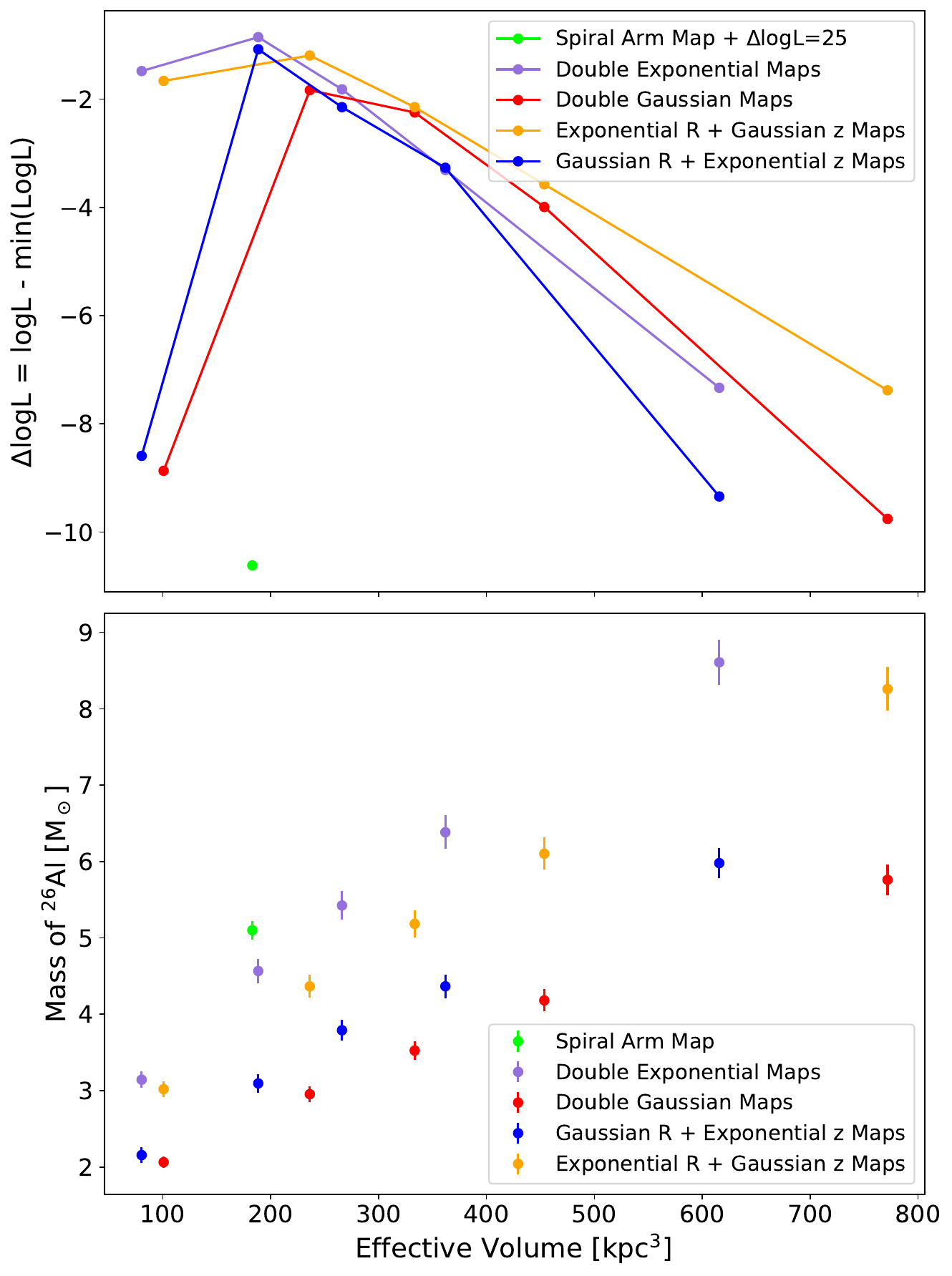}
    \caption{Top: Log-likelihood of different global disk models as a function of effective volume. Note that the effective volume is given in Tab.\,\ref{models}, so that different combinations of $R_E$ and $h_E$ can lead to the same volume but different likelihoods. Due to computational constraints, we use $V_{\rm eff}$ as a summary variable. All models show minimal log-likelihood for \(R_E=5.0\,\) kpc and \(h_E=0.6\,\)kpc. \\
    Bottom: Corresponding masses for the different models. For the best fit, the model suggestions lay between  $(2.95\pm0.10)$\,M$_\odot$ and $(4.57\pm0.16)$\,M$_\odot$ within 15\,kpc Galactocentric radius.}
    \label{loglik}
\end{figure}

\begin{figure*}[!ht]
    \sidecaption
    \includegraphics[width=12cm]{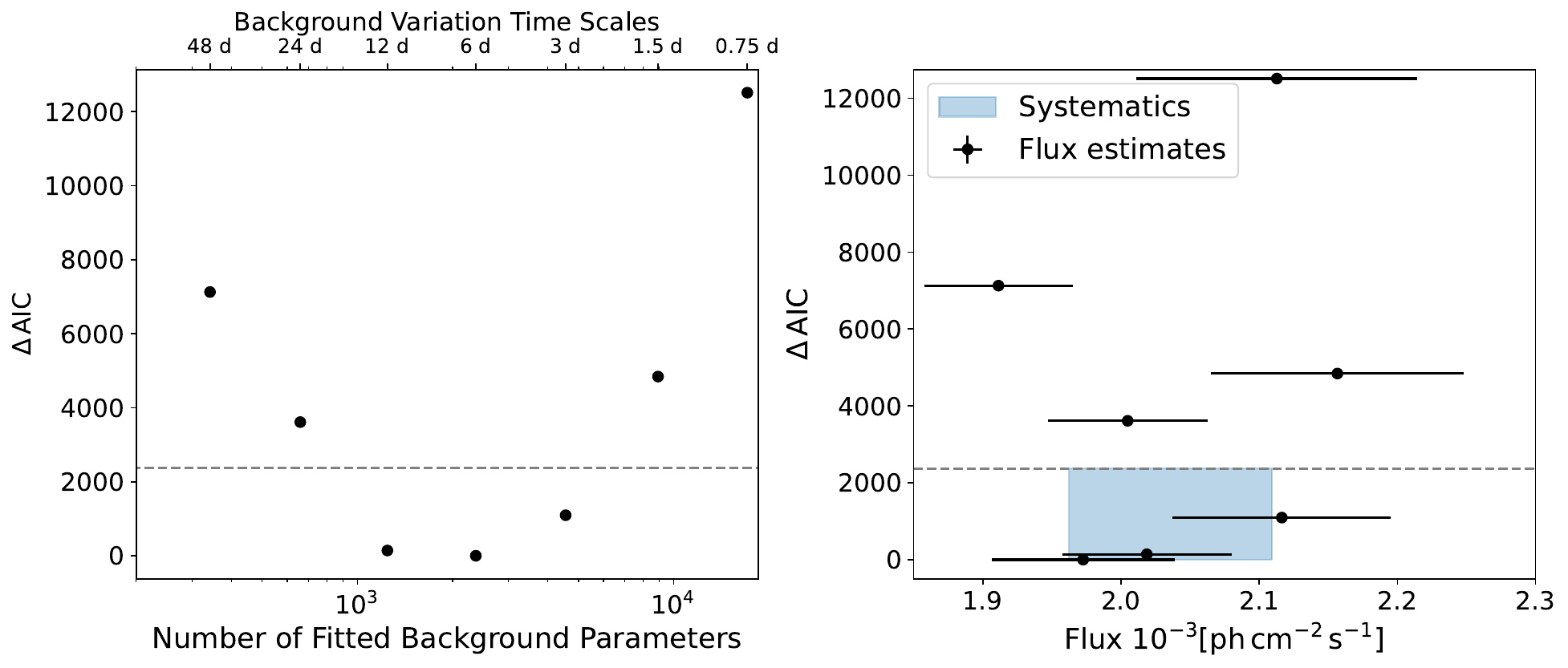}
    \caption{{Background variability and systematics of the Data Set. Left: Scan for optimal Background Variation Time Scale as indicated by the minimum AIC value. For an adequate fit, renormalization every two INTEGRAL orbits is sufficient. Right:Corresponding flux estimates. The standard deviation of the fluxes whose $\Delta$AIC is below the number of fitted parameters at optimum AIC is used to estimate the systematics.}}
    \label{fig:AIC_quer}
\end{figure*}

The GCE predictions depend on both observationally constrained quantities and theoretical model inputs.
The Galactic nova rate is constrained by the observed nova frequency, while the adopted delay-time distribution follows from binary stellar-evolution modelling, and the CO/ONe nova fraction is chosen consistently with previous GCE studies.
The nucleosynthesis yields, by contrast, are theoretical predictions rather than quantities calibrated directly to the Galactic \nuc{Al}{26} observations.
\citet{vasini2024tracingmilkywayspiral} adopt the nova yields of \citet{Jose2007}, averaged over the considered white-dwarf masses and combined assuming a 70\% CO and 30\% ONe nova population.

The absolute yield normalisation is almost irrelevant here because we compare the observed absolute radial profile to the relative spatial distributions predicted for the different source populations.
Metallicities and stellar yields are already treated consistently in the GCE calculations, and additional enrichment from slowly evolving bulge novae is included in the innermost Galactic regions following \citet{vasini2024tracingmilkywayspiral}.
Consequently, our comparison tests whether the predicted spatial distributions are compatible with the observed \nuc{Al}{26} profile rather than the absolute nucleosynthetic yields.
Theoretical nova calculations span a wide range of \nuc{Al}{26} yields and ejecta properties, depending on quantities such as white dwarf mass and composition, accretion history, mixing assumptions, and the adopted nuclear reaction network.
For example, the theoretical \nuc{Al}{26} yields discussed before span several orders of magnitude.
The GCE profiles used here adopt one specific nova-yield prescription.
We do not rerun the GCE calculation with alternative nova-explosion yield or ejecta-mass grids.
Such alternatives can, in principle, be incorporated by replacing the adopted yield prescription and recalculating the chemical-evolution model.
A change that only rescales the nova yield uniformly would largely be absorbed by the free nova normalisation in our fit, whereas changes that modify the relative radial source profile could affect the inferred source decomposition.
These theoretical GCE-profile uncertainties are not propagated in the present analysis, and the inferred source fractions should therefore be understood as conditional on the adopted GCE profiles.

As an alternative source, the winds of very massive stars \citep[VMS;][]{Higgins2023_VMS}, i.e. stars with initial masses between 150 and 300\,M$_\odot$, might also be considered as a contributor to the global \nuc{Al}{26} budget in the Galaxy. \citet{Martinet_2022} studied the output of rotating and non-rotating VMS models for different initial masses and metallicities. They used GENEC \citep{GENEC_Eggenberger} to compute the models of different initial mass and metallicity conditions. They chose reaction rates of $\nuc{Mg}{25}(p,\gamma)\nuc{Al}{26}^ {g,m}$ \citep[][treating the isomeric and the ground state of \nuc{Al}{26} as different populations]{Iliadis_2001}, $\nuc{Al}{26}(p,\gamma)\nuc{Si}{27}$ \citep{Champagne_1993_reactionrates}, $\nuc{Al}{26}(n,\alpha)\nuc{Na}{23}$, and $\nuc{Al}{26}(n,p)\nuc{Mg}{26}$ \citep{Caughlan_1988_reactionrates}, to obtain their mass estimates. 
For the mass-loss rates of massive stars, the radiative losses were taken from \citet{Vink_2001} and \citet{deJager_1988}, while the mass loss rates for the Wolf-Rayet phase are from \citet{Graefener_2008} and \citet{Nugis_2000}. They apply a \citet{Salpeter_1955} initial mass function (IMF) and calculate the \nuc{Al}{26} yields throughout the Galaxy. They present results for four different sets of stellar population, comparing rotating and non-rotating models with and without VMS. They set the maximum mass for VMS to $300\,\text{M}_\odot$, based on observations of VMS in the Tarantula nebula \citep{Bestenlehner_2020, Brands_2022}, and find that rotating models in general produce more \nuc{Al}{26} for both massive and very massive populations \citep[see also][]{Limongi_Chieffi_18, Falla_2025_rotRatio}. Taking into account VMS increases the Galactic budget of \nuc{Al}{26} by $120$--$150\%$.  Combining rotation and VMS in their model, they find the production rate of \nuc{Al}{26} to be four times larger than for models accounting neither. By pushing the upper mass limit to $500\,\text{M}_\odot$, they obtain an even higher \nuc{Al}{26} output, increasing the results for lower mass VMS by a factor of two.

\citet{Martinet_2022} show that even a small population of VMS can impact the Galactic amount of \nuc{Al}{26} drastically and might overall ought to be taken into account in nucleosynthesis research. However, the geometric distribution of VMS would be similar to massive stars, so that our approach, which treats only the {relative spatial} distribution in radial bins, would hardly be affected.

{ The massive-star model underproduces the total observed \nuc{Al}{26} inventory by up to a factor of four.}
\begin{figure*}[t!]
  \subfloat[][]{\includegraphics[width=.3\linewidth]{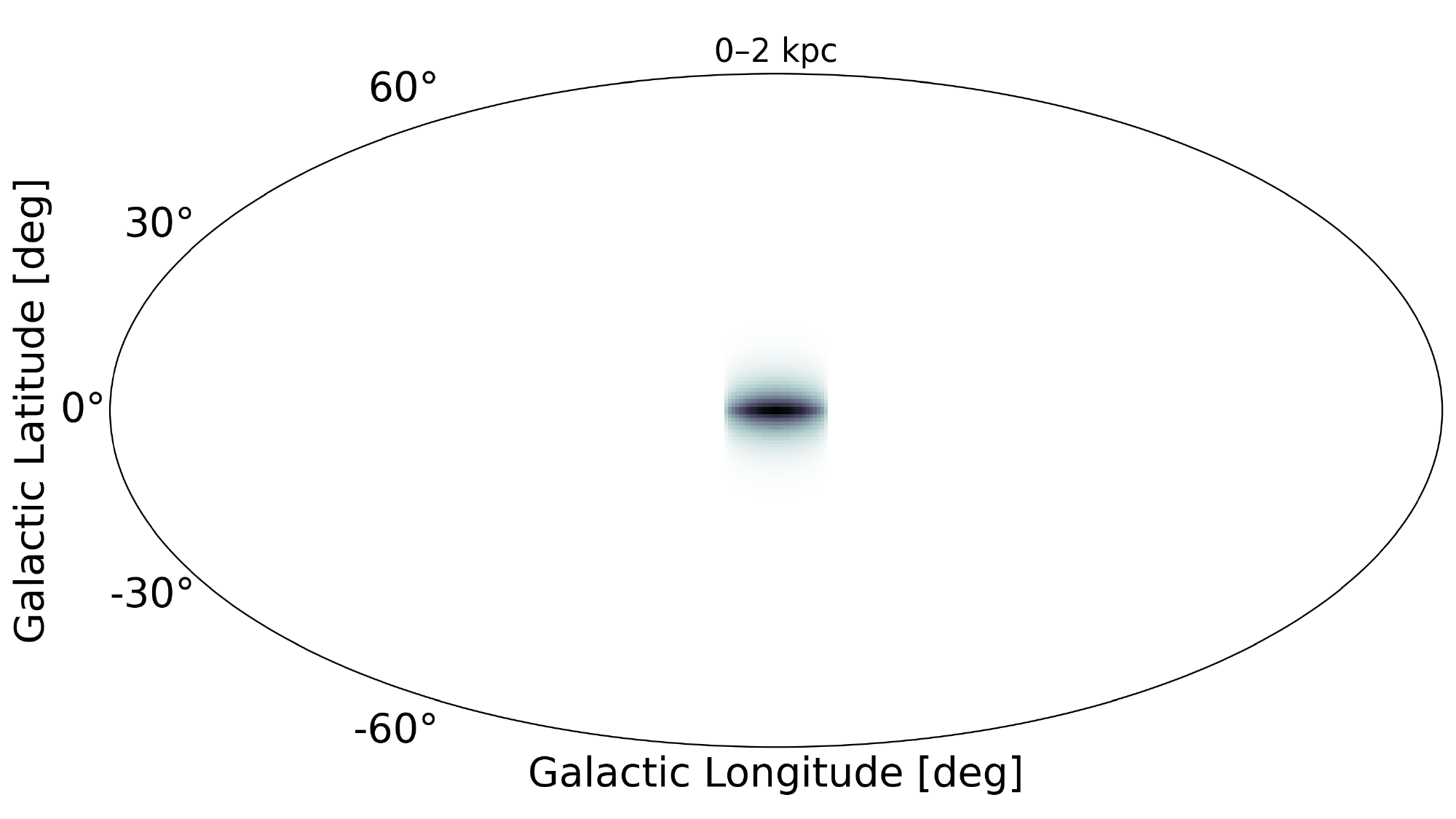}} 
      \quad
     \subfloat[][]{\includegraphics[width=.3\linewidth]{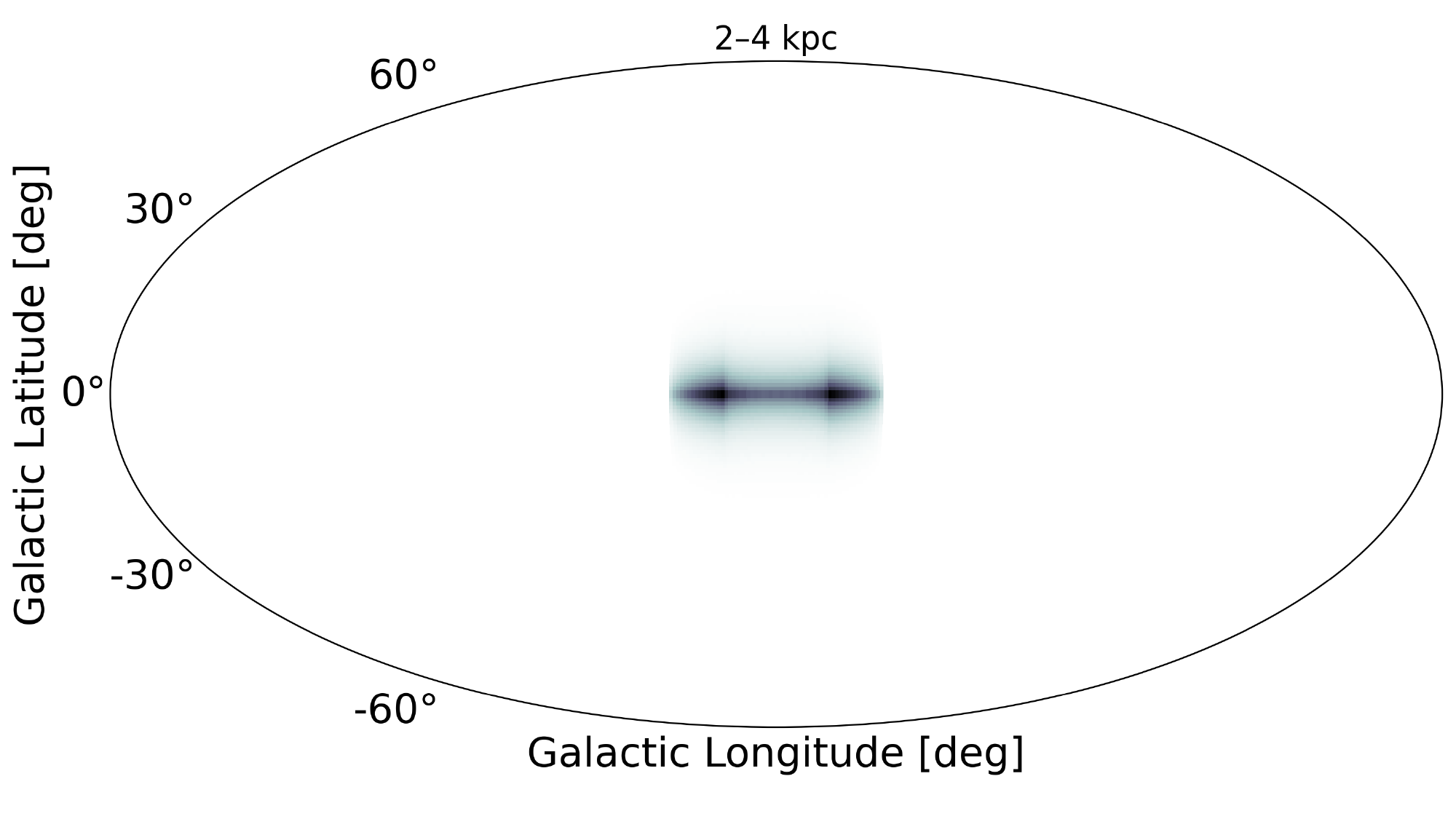}}
     \quad
     \subfloat[][]{\includegraphics[width=.3\linewidth]{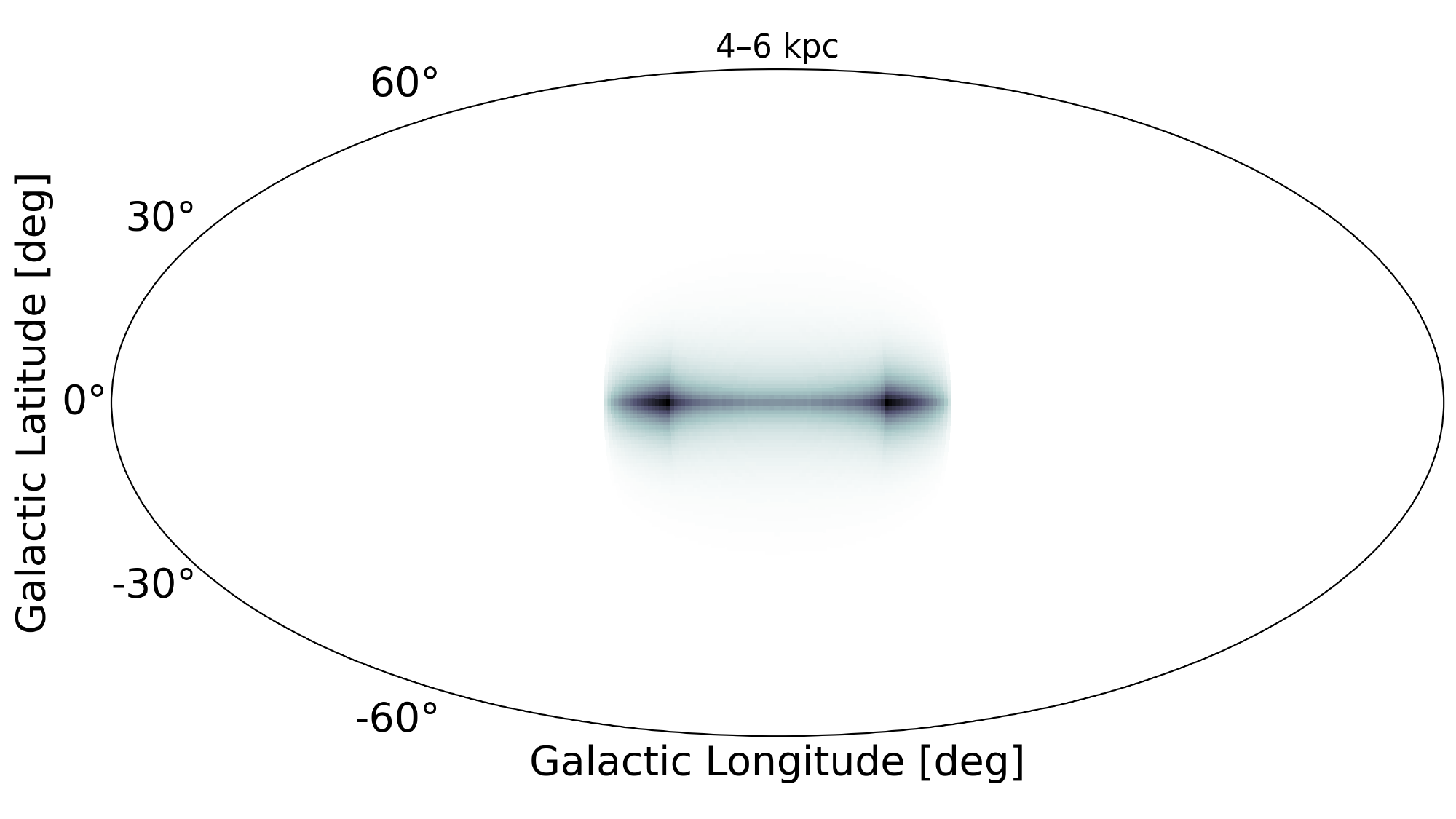}}
      \quad
     \subfloat[][]{\includegraphics[width=.3\linewidth]{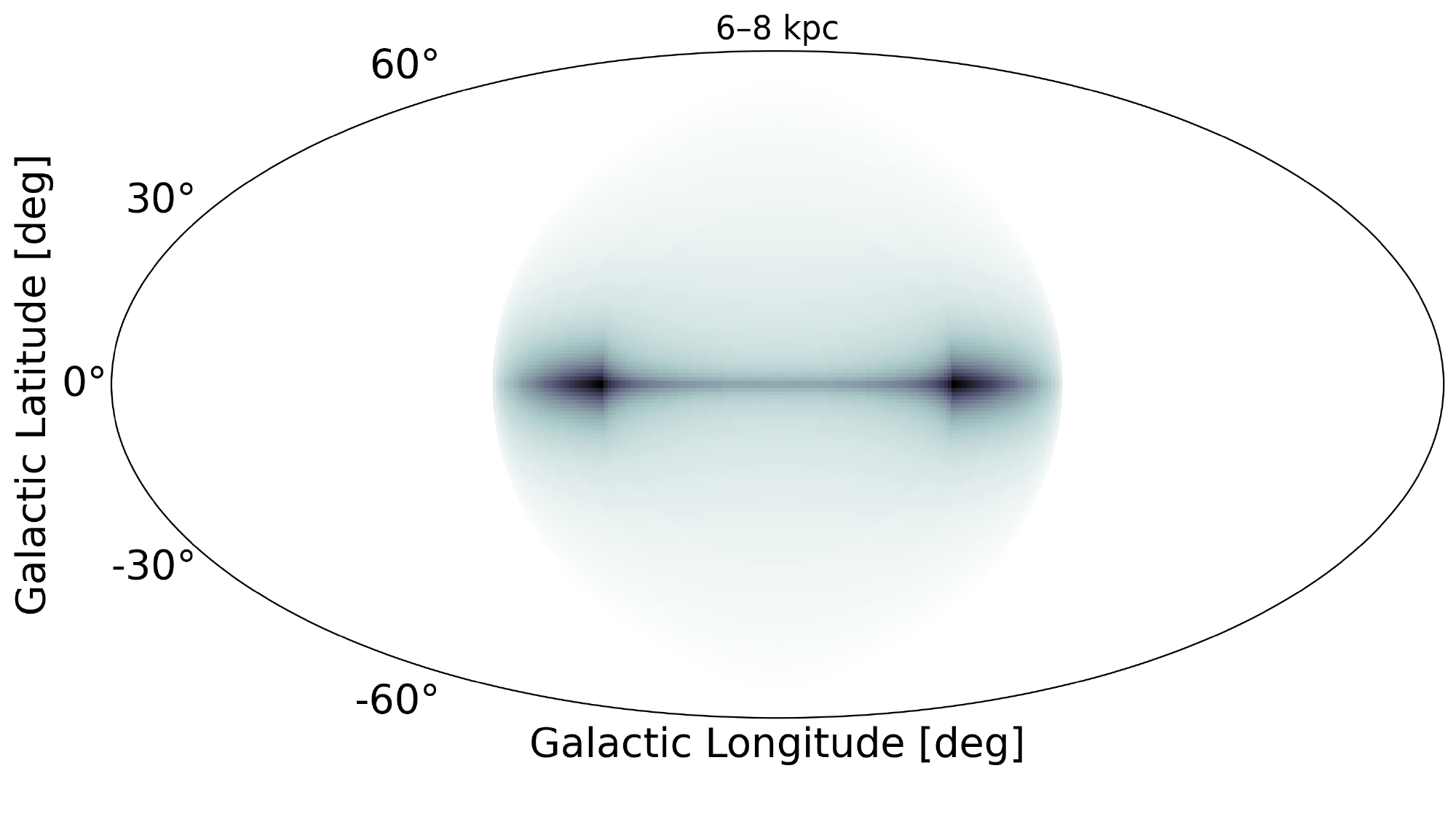}}
     \quad
      \subfloat[][]{\includegraphics[width=.3\linewidth]{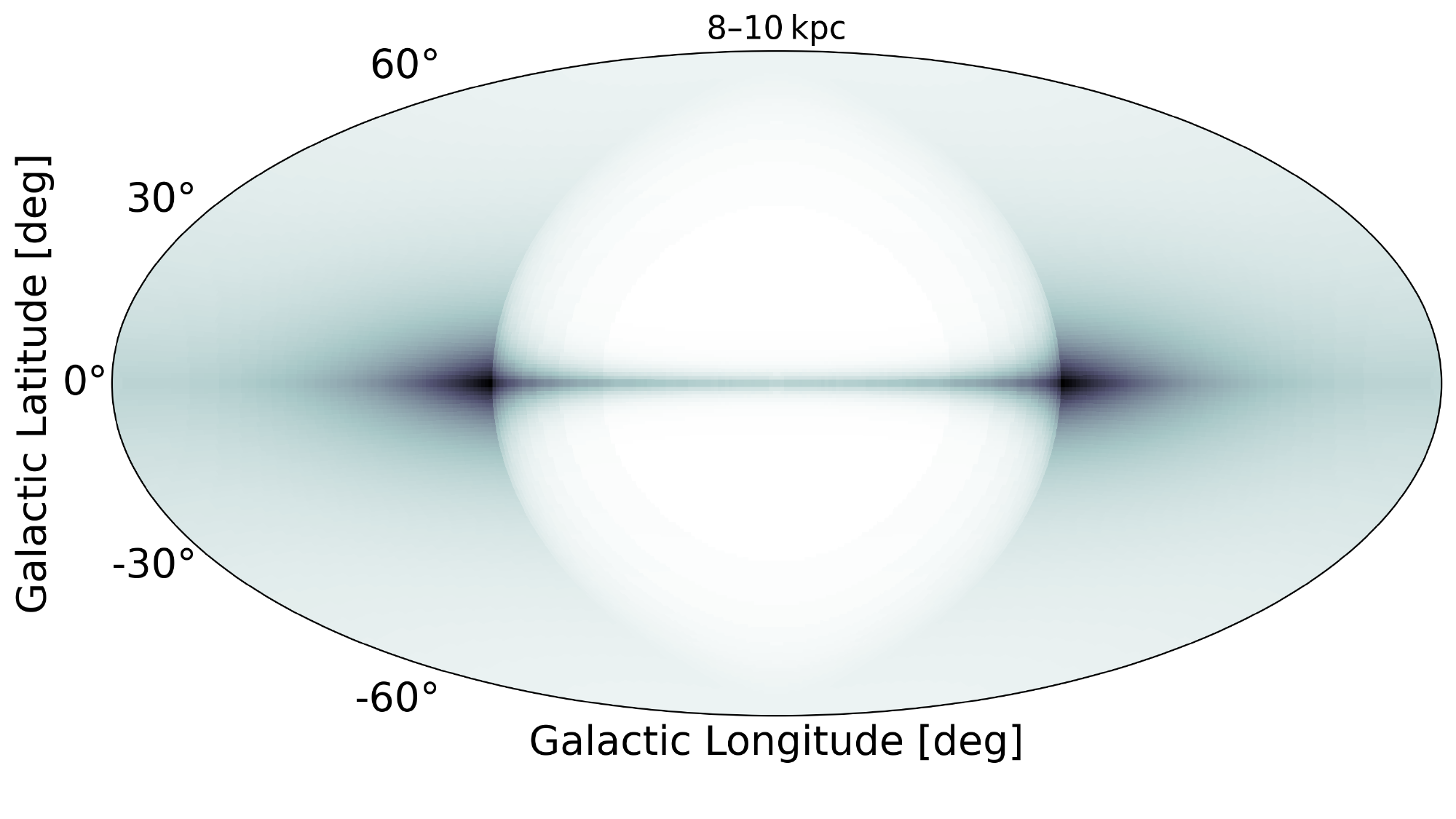}}
     \quad
      \subfloat[][]{\includegraphics[width=.3\linewidth]{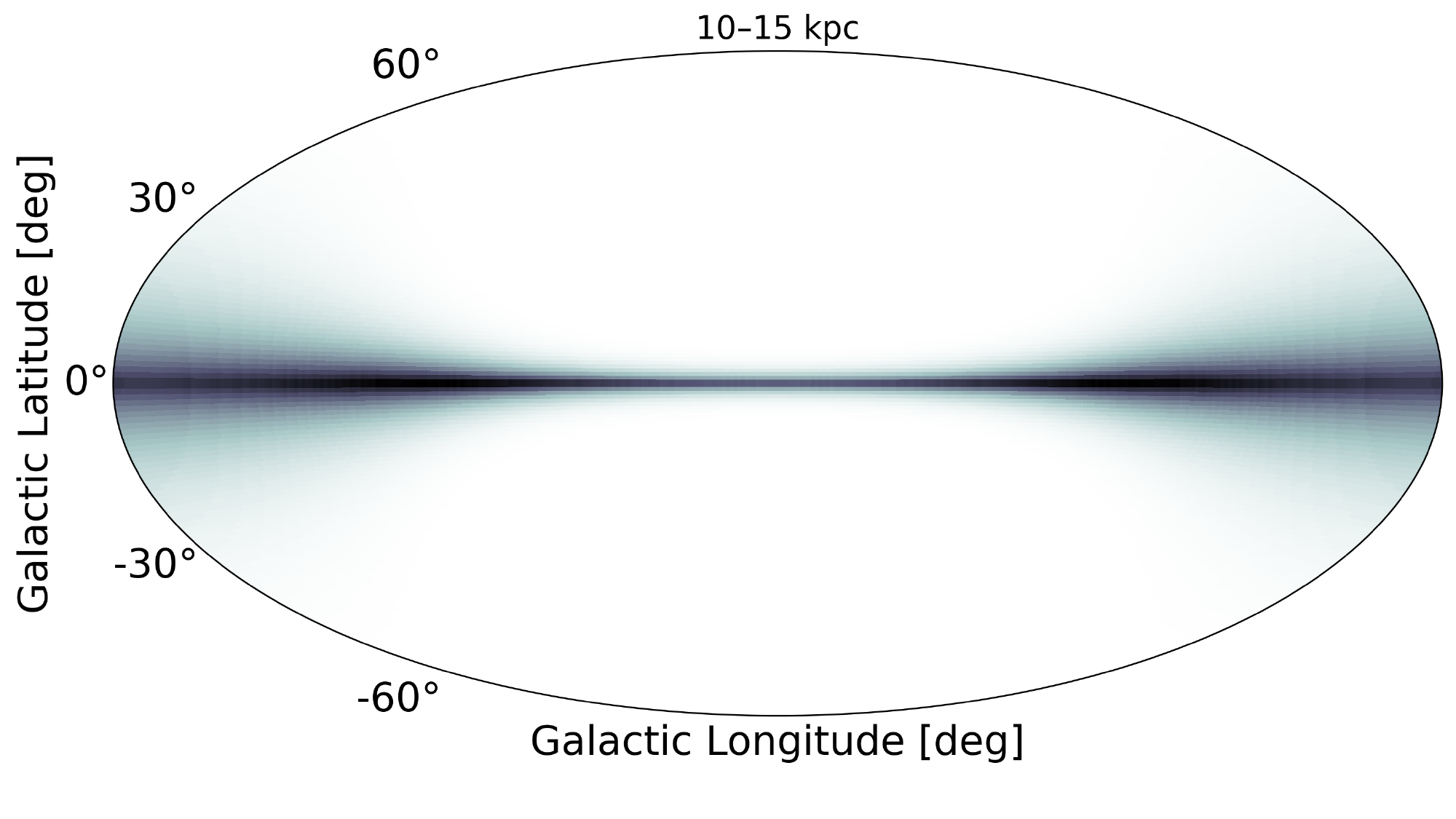}}
     \quad
     \caption{Model maps, created by line-of-sight integration of a double exponential ansatz, Table\,(\ref{models}). In this example, scale values of $R_E=5.0$\,kpc and $h_E=0.6$\,kpc were applied. The flux is given in arbitrary units from yellow, maximum flux, to black, zero flux. Here, the Galaxy was cut into rings from 0–2 kpc, 2–4 kpc, 4–6 kpc, 6–8 kpc, 8–10 kpc, and 10–15 kpc of Galactocentric distance.}
     \label{ModelMaps_fitted}
\end{figure*}

\section{INTEGRAL/SPI dataset and analysis}\label{INTEGRAL/SPI_analysis}
\subsection{Dataset}\label{sec:dataset}
We are using 20.5\,yr of INTEGRAL/SPI observations of the entire sky.
We select the energy range between 1790--1840\,keV and bin the data in 0.5\,keV bins for the initial construction of the background model, and then rebin the data into three bins of 15\,keV, 8\,keV and 27\,keV, respectively, for the final analysis.
In total, we select 143,833 pointed observations (``pointings'') with an average observation time (dead-time corrected) of 2128\,s.
We filter out 473 pointings because no self-consistent background model could be created or the residuals in these observations were too large.
The final dataset therefore has 305\,Ms of on-time across the sky.
We note that the exposure of INTEGRAL/SPI is very patchy with long exposure of the Galatic bulge and plane but hardly any observations at higher latitudes $|b| \gtrsim 15^\circ$, except for the regions around Virgo, the LMC or Orion-Eridanus.
The scattered observations are properly taken into account by our methodology and since most observations are found around latitudes $b \sim 0^\circ$, the performed analysis is barely affected by systematics from an observational bias.

\subsection{Data Analysis}\label{sec:data_analysis}
Raw SPI data is modeled by two general components in a linear combination: instrumental background and astrophysical sources.
The forward model reads
\begin{equation}
    m_{dp}(\mathbf{\theta}) = \sum_{i=1}^{N_S} \theta_i R_{dp}^{\ell b} \otimes M_{i,\ell b} + \sum_{j=1}^{N_B}\theta_j B_{j,dp}\mathrm{,}
    \label{eq:spimodfit}
\end{equation}
where $m_{dp}(\mathbf{\theta})$ are the expected model counts in detector $d$ and pointing $p$, depending on the parameters $\mathbf{\theta}$ that scale the $N_S$ sky models and $N_B$ background models in a maximum likelihood fit using the Poisson statistics within OSA/\texttt{spimodfit} \citep{Courvoisier2003_ISDC,Strong2005_spimodfit}.
Each sky model $M_{i,\ell b}$ that is pixelated on a grid of Galactic coordinates $\ell$ and $b$ is transformed into the data space through the imaging response function $R_{dp}^{\ell b}$.
The difference between sky models and background models, $B_{j, dp}$, is that the background does not depend on the observing direction of the instrument.
The background is modeled self-consistently according to \citet{Siegert2019_Background} and \citet{Diehl2018_Background} who built a spectral response and background database over the mission duration.
We find that for an adequate fit, a background re-normalization time scale of one parameter every two INTEGRAL orbits ($\sim$6\,d) for each component, instrumental background continuum and instrumental background lines, is sufficient.
The total numbers of fitted background parameters is then $n_{\mathrm{par}} = 2368$.
We estimate a systematic uncertainty on the total flux estimates by investigating still-adequate numbers of background parameters.
In Fig.\,\ref{fig:AIC_quer}, we show Akaike Information Criterion (AIC) compared to the best-fit model ($\Delta\mathrm{AIC} = 0$) as a function of the number of fitted background parameters (left) and the resulting \nuc{Al}{26} 1809\,keV flux (right).
Based on the fact that $n_{\mathrm{par}} = 2368$ provides the minimum $\mathrm{AIC}$, we consider $\Delta\mathrm{AIC}$-values of up to $2368$ to describe the dataset still well \citep{Siegert2019_Background}.
From the variance of the resulting flux values, between $1.96$ and $2.12 \times 10^{-3}\,\mathrm{ph\,cm^{-2}\,s^{-1}}$, we derive a systematic uncertainty coming from different background variability time scales of $\sigma_\mathrm{syst} = 1.4 \times 10^{-4}\,\mathrm{ph\,cm^{-2}\,s^{-1}}$, or, equivalently, 7\%.
This value should be considered for all extracted flux values and, consequently mass and mass fraction estimates.
The number of parameters to describe the sky distribution is then explained in the following sections.

\section{modelling the 1.8\,MeV sky -- Galactic disk models}\label{galactic_disk_models}
We aim to estimate the persistent mass of \nuc{Al}{26} in the Milky Way as a whole and then as a function of Galactocentric radius.
Our strategy is to determine, first, a set of shape parameters for the Galactic-wide fit using four different models that had been found in previous studies to describe the \nuc{Al}{26} sky well \citep[e.g.,][]{Oberlack1996_COMPTEL26Al,Knoedlseder1999_COMPTEL26Al,plueschke2001comptel1809mevsurvey,Diehl2006_26Al,dissertation_Pleintinger}.
The total flux of the 1.8\,MeV line at the best-fit values of the shape parameters then converts to a persistent \nuc{Al}{26} mass in the Milky Way.
In the second step, the parameters $R_E$ and $h_E$ are fixed for each emissivity model assumption, and the total model is split up into radial bins (rings) and fitted simultaneously.
This means any variation from the initial radial assumption (exponential or Gaussian) is re-determined on smaller scales.
Thus, each component obtains a fitted 1.8\,MeV flux which converts to an \nuc{Al}{26} mass inside that ring.
In the third step, the resulting masses per radial bin are used to estimate the contributions of classical novae and (very) massive stars to Galactic \nuc{Al}{26} by comparison to different GCE models.

\subsection{Galactic-wide emission}\label{sec:MW_models}
In total, we use four different models to create three-dimensional emissivity distributions, $\rho(x,y,z)$, and create full-sky maps, $F(\ell,b)$, as a function of Galactic longitude $\ell$ and latitude $b$, by line-of-sight integration.
The emissivities we use are summarized in Tab.\,\ref{models}.
The flux maps are calculated by
\begin{equation}
    F(\ell,b) = \frac{1}{4\,\pi} \int_0^\infty\,\mathrm{d}s\,\rho\left(\Tilde{x}(s),\Tilde{y}(s),\Tilde{z}(s)\right)\mathrm{,}
    \label{eq:los}
\end{equation}
where the line of sight is defined as
\begin{align}
    \Tilde{x}(s,\ell,b) & =  x_\odot + s\,\cos \ell\,\cos b\,, \nonumber\\
    \Tilde{y}(s,\ell,b) & =  y_\odot + s\,\sin \ell\,\cos b\,, \nonumber\\
    \Tilde{z}(s,\ell,b) & =  z_\odot + s\,\sin b\,,
    \label{eq:los_coordinates}
\end{align}
with $\left(x,y,z\right)_\odot$ being the coordinates of the Sun as the position of the observer, and $s$ being the line-of-sight distance.
The normalisation is obtained by calculating the ``effective volume'' $V_\mathrm{eff}$ of the emissivities,
\begin{equation}
    V_\mathrm{eff} = \int\mathrm{d}\Omega\,\int_0^\infty\mathrm{d}s\,s^2\,\frac{\rho\left(\Tilde{x}(s),\Tilde{y}(s),\Tilde{z}(s)\right)}{\rho_0}\mathrm{,}
    \label{eq:effective_volume}
\end{equation}
so that, finally, the luminosity, $L_0$, relates the effective volume to the emissivity amplitude,
\begin{equation}
    \rho_{0} \equiv \frac{L_{0}}{V_\mathrm{eff}}\mathrm{.}
    \label{eq:emissivity_norm}
\end{equation}
The effective volume of an emission model describes the extent of the emitting region:
We use $V_\mathrm{eff}$ as a measure of the spatial weight of a given emission model.
It is the volume over which a uniform emissivity $\rho_0$ would produce the same total luminosity as the actual, non-uniform model.
Thus, $V_\mathrm{eff}$ accounts not only for the size of the emitting region, but also for how the emission is distributed within it.

The tested models include a doubly-exponential disk, a doubly-Gaussian disk, as well as mixed configurations, all with the two parameters of scale height $h_E$ and scale radius $R_E$ (see Tab.\,\ref{models}).
Because testing different combinations in the $(R_E,h_E)$-plane with such a large data set for different models is computationally very time consuming, we select the combined parameter of effective volume to determine adequate values of $R_E$ and $h_E$.

\begin{table*}[t!]
\caption{Density profile and effective volume for different Galactic models of the \nuc{Al}{26} emissivity.}                 
\label{models}    
\centering                        
\begin{tabular}{c c c c c}      
\hline\hline               
& Disk model & Density & Effective Volume \\         
\hline                      
  Case 1 &   Double exponential & $\rho(x,y,z)=\frac{L_0}{4\pi R_E^2h_E}\exp\left(-\frac{|z|}{h_E}\right)\exp{\left(-\frac{-R_{x,y}}{R_E}\right)}$ & $ 4\pi R_E^2h_E $\\    
  Case 2 & Double Gaussian & $\rho(x,y,z)=\frac{L_0}{(2\pi)^{3/2} R_E^2h_E}\exp\left(-\frac{z^2}{2h_E^2}\right)\exp{\left(-\frac{R^2_{x,y}}{2R_E^2}\right)}$ & $(2\pi)^{3/2} R_E^2h_E$ \\
 Case 3 & Exponential scale radius, Gaussian scale height & $\rho(x,y,z)=\frac{L_0}{(2\pi)^{3/2} R_E^2h_E}\exp\left(-\frac{z^2}{2h_E^2}\right)\exp\left(-\frac{R_{x,y}}{R_E}\right)$ & $(2\pi)^{3/2} R_E^2h_E$ \\
 Case 4&  Gaussian scale radius, exponential scale height &  $\rho(x,y,z)=\frac{L_0}{4\pi R_E^2h_E}\exp\left(-\frac{|z|}{h_E}\right)\exp{\left(-\frac{R^2_{x,y}}{2R_E^2}\right)}$ & $4\pi R_E^2h_E $\\
 Case 5 & NE 2001 & \\
\hline                                  
\end{tabular}
\end{table*}

The emissivity profiles are listed in Tab.\,\ref{models}, where $\rho(x,y,z)$ is the spatial emissivity of the 1.8\,MeV line in units of $\mathrm{cm^{-3}\,s^{-1}}$, given a luminosity,
\begin{equation}
L_0 = \frac{M_{26}p_{26}}{m_{26}\tau_{26}}\mathrm{,}
\label{eq:lumi}
\end{equation}
with the persistent \nuc{Al}{26} mass $M_{26}$, and the isotope properties $p_{26} = 0.997$ (probability to emit a 1.8\,MeV photon after decay), $m_{26} = 26\,\mathrm{u}$ (isotope mass), and $\tau_{26} = 1.04$\,Myr (decay time).
In Fig.\,\ref{SkyMap}, we show the diffuse 1.8\,MeV emission overlaid with exemplary outlines of the models described above.  The levels are chosen to enclose 90\,\% of the total emission.

The distribution of \nuc{Al}{26} throughout the Galaxy depends on its parent source population. Fig.\,\ref{Vasinivsus} shows the distribution of \nuc{Al}{26} as predicted for a doubly-exponential and a doubly-Gaussian disk with \(R_E =5.\,8\,\text{kpc}\), \(h_E=0.\,8\,\text{kpc}\) according to \citet{Combinations}, and the resulting persistent \nuc{Al}{26} mass of 4.5\,M$_\odot$ in total (equivalent to 3.8\,M$_\odot$ up to 15\,kpc), compared to the previously discussed GCE model of \citet{vasini2024tracingmilkywayspiral}.
As an additional, more sophisticated spatial model, we use the electron density distribution NE2001 from \citet{NE2001_paper} as it was shown to represent the \nuc{Al}{26} morphology adequately \citep{Knoedlseder1999_COMPTEL26Al,dissertation_Pleintinger}. Here, we use the components of a thick disk, thin disk, and spiral arms, and leave out the nearby components of the ISM, Galactic centre, clumps and voids as they put too much emphasis on specific coordinates in the 1.8\,MeV map. The resulting equivalent persistent mass is then 3.6\,M$_\odot$.
We note that while the masses can be significantly different, the total flux of the Galaxy at 1.8\,MeV would be very similar, changing by less than 1\% for any spatial model.
While for novae the mass peak is found between \(4\,-5\,\text{kpc}\), the distribution of massive stars peaks between \(6\, - 7\, \text{kpc}\). In addition, the rise from the Galactic centre is steeper for novae compared to massive stars. Both, a doubly-exponential and a double-Gaussian disk with these parameters would peak between 5 and 6\,kpc. 

The choice of the disk parameters is not unique, so that we use a set of different combinations of $R_E$ and $h_E$ to estimate a systematic uncertainty in addition to the statistical uncertainties from the maximum likelihood fit. \citet{Combinations} present good fit value combinations for a doubly-exponential disk model. Using their results, we probed five different effective volumes and fitted the resulting global maps to our data set. This provides an adequate starting point for selecting models when the Galaxy is split into multiple components (Sec.\,\ref{sec:26Al_rings}).

We judge the relative quality of the fits by their log-likelihood and calculate the corresponding mass for each of them. Both are shown in Fig.\,\ref{loglik}. While, in principle, different combinations of $R_E$ and $h_E$ can give the same values of $V_{\rm eff}$ (see Tab.\,\ref{models}), we only use a restricted amount of tested $V_{\rm eff}$ values as a summary variable due to computational constraints of testing thousands of different combinations. We note that by using the selected values, a large range of possible Galactic morphology shapes are already covered, so that {the resulting variance of these combinations serves as an estimate of the systematic uncertainty associated with the adopted SPI emission morphology}, which is equal or smaller than the statistical one (see Fig.\,\ref{MassModels}). {This does not include uncertainties in the GCE source profiles used for the subsequent source decomposition.} The parameters from \citet{Combinations}, $(R_E,h_E) = (5.8 \pm 0.6,0.77 \pm 0.17)$\,kpc for the doubly-exponential disk, corresponding to $V_{\rm eff} = (326 \pm 102)\,\mathrm{kpc}^3$, were found as best-fit values, so that based on that, we test a similar range for our emissivity profiles {for Cases 1--4.}
Given the likelihood values in Fig.\,\ref{loglik}, top, for Cases\,1\&3, effective volumes from 100 to 290\,kpc$^3$ provide adequate fits. For Cases\,2\&4, the range would be 180 to 350\,kpc$^3$. The resulting \nuc{Al}{26} mass values are found in the range  3.0--5.5\,M$_\odot$.
All global models show their best fit value for $R_E = 5.0$\,kpc and $h_E=0.6$\,kpc, corresponding to $V_\text{eff}=188$\,kpc$^3$ (Case\,1 and 4) and 236\,kpc$^3$  (Case\,2 and 3), and the corresponding best fit mass values fall between 
$3.0$\,M$_\odot$ and $4.6$\,M$_\odot$ for the total Galaxy. {For a global spatial model, the Spiral Arm model (Case 5), with $V_\text{eff}=182$\,kpc$^3$, is, compared to the best fit, rather bad to describe the global distribution of \nuc{Al}{26}.}
Considering the mass-to-flux scaling as described above, the flux in the range up to 15\,kpc and up to the entire scale of the Milky Way would be almost identical even if the masses are set to zero beyond 15\,kpc.
This means the masses derived for the entire Galaxy should always be considered upper limits for which 10--20\% could systematically be removed.

\subsection{Galactocentric \nuc{Al}{26} profile}\label{sec:26Al_rings}
\begin{figure}[t]
    \centering
    \includegraphics[width=\linewidth]{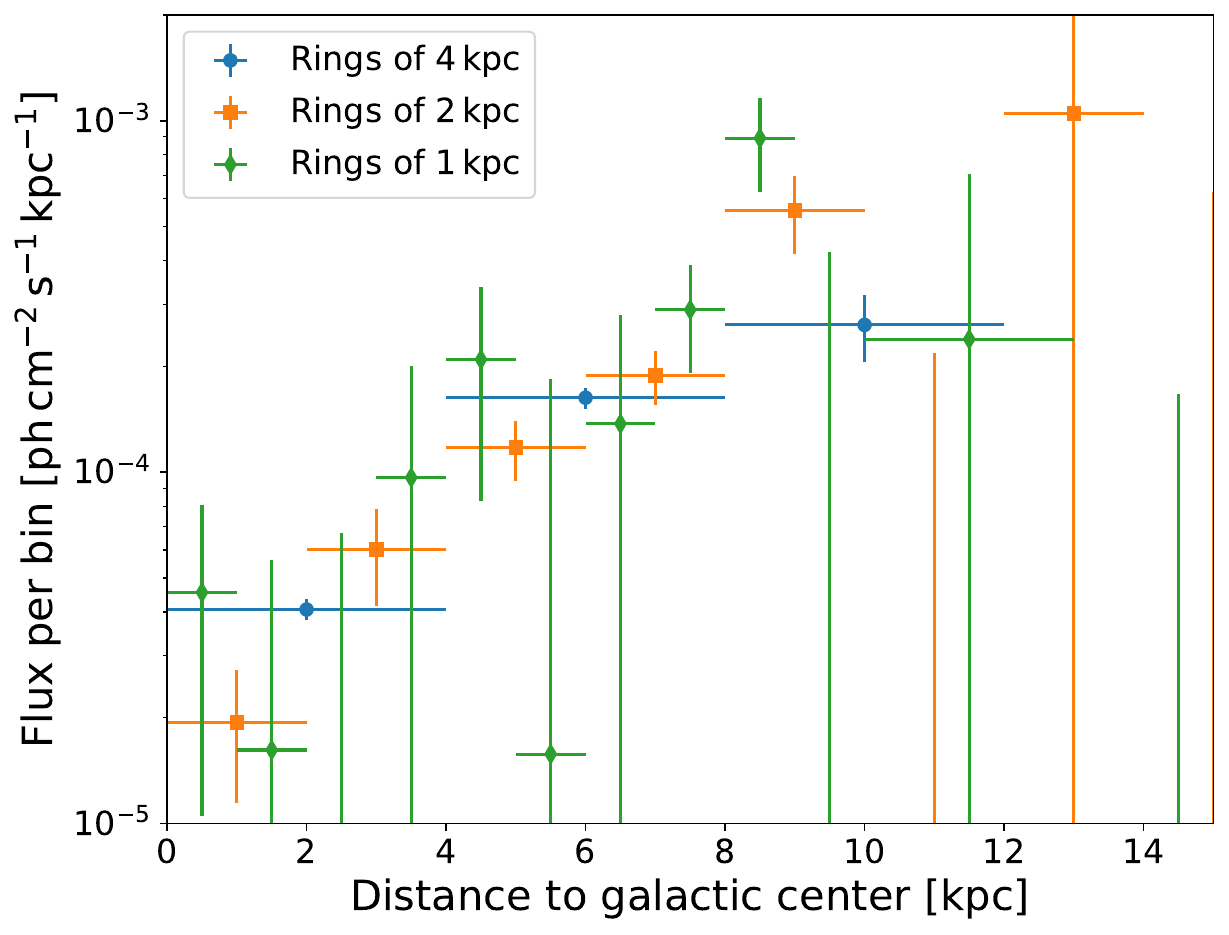}
    \caption{Measured flux distribution of the \(1.8\,\text{MeV}\) line as a function of Galactocentric radius for a double exponential disk model with scale parameters $R_E=5.0\,$kpc and $h_E=0.6\,$kpc with varying ring size.}
    \label{fluxes}
\end{figure}

\begin{figure}[h]
    \centering
    \includegraphics[width=\linewidth]{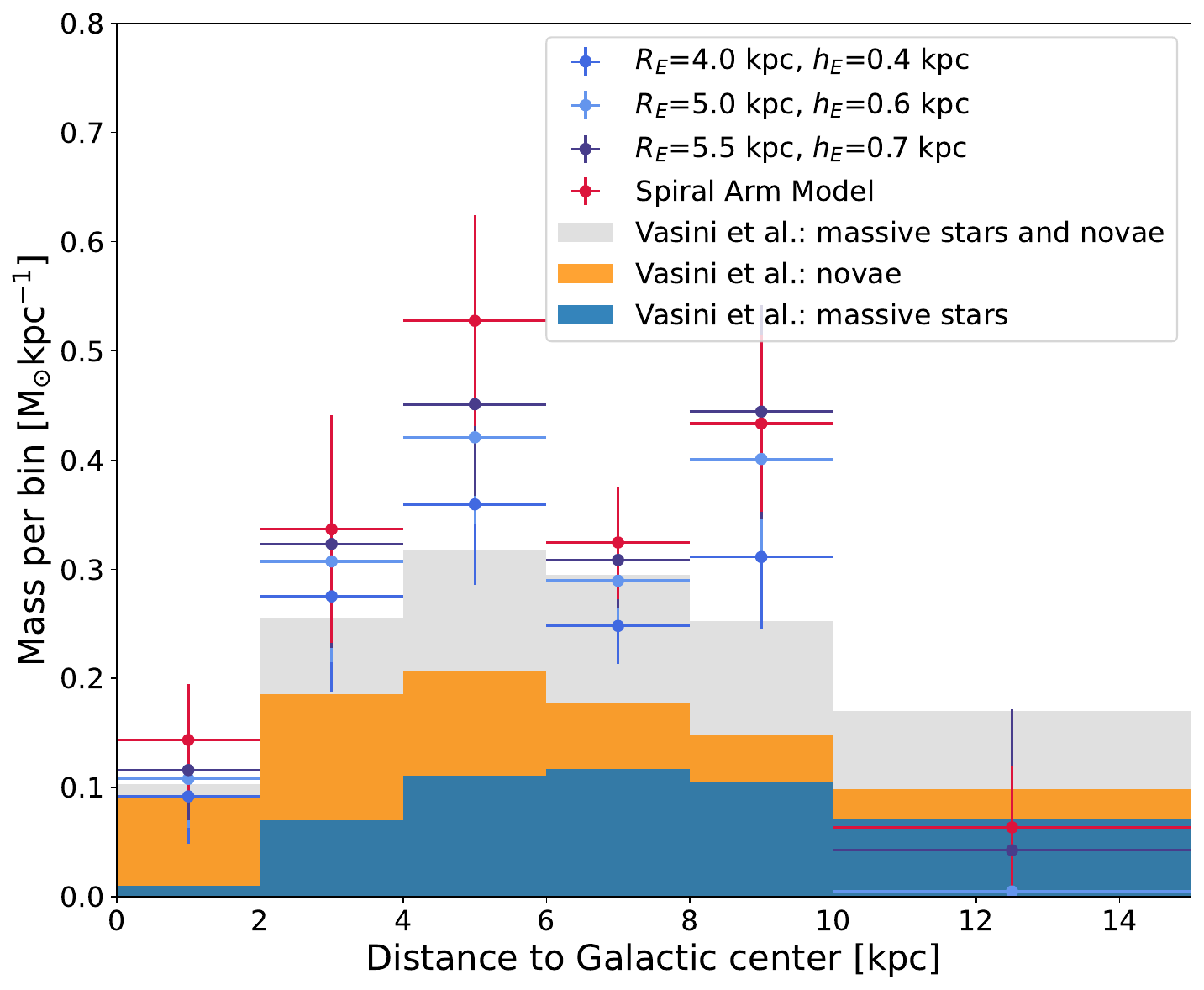}
    \caption{\nuc{Al}{26} mass distribution for different scale value combinations of a double exponential model {and the NE2001 Spiral Arm model}. The masses per ring are consistent with each other, within uncertainties. In grey, the results of \citet{vasini2024tracingmilkywayspiral} for both, massive stars and novae are shown. The source composition of those values is indicated for each bin. For the 8--10\,kpc ring the model expectation does not align with measurements within the respective uncertainties.}
    \label{MassModels}
\end{figure}

A single disk model falls short in explaining the combined GCE model, for which we separate it into rings from 0--2\,kpc, 2--4\,kpc, 4--6\,kpc, 6--8\,kpc, 8--10\,kpc, and 10--15\,kpc, and allow these components to be fitted simultaneously. We explain the choice of these radial bin sizes in the following. This means that instead of one map (Fig.\,\ref{SkyMap}), six maps (Fig.\,\ref{ModelMaps_fitted}) are used to fit the raw INTEGRAL/SPI data. 
{Fig.\,\ref{ModelMaps_fitted} shows the projected geometrical templates for Galactocentric rings assuming a doubly-exponential disk with scale radius $R_E=5.0$\,kpc and scale height $h_E=0.6$\,kpc.
Because of the analytical forms, we can convert the resulting fluxes from each concentric ring into an estimate of the persistent \nuc{Al}{26} mass in that ring by Eqs.\,(\ref{eq:los})--(\ref{eq:lumi}).
We note that this is not an independent measurement of the \nuc{Al}{26} mass in those rings but model dependent.
Given that different emission templates lead to largely similar masses (see Figs.\,\ref{fluxes} \& \ref{MassModels} in the following), we deem this procedure a robust mass estimate.}
By comparing the peaks and rise, we aim to calculate the overall contribution of novae and massive stars after finding the best fit.

In order to describe the 1.8\,MeV emission in more detail, we repeatedly cut up the Galactic emissivity profiles into concentric rings around the Galactic center and perform line-of-sight integrations to obtain ring-like maps. We cut small rings in each iteration to obtain a reasonable spacing of the rings while at the same time not losing too much information from increased uncertainties when more components are added. First, we cut up the Galaxy into two parts, with a width of \(8\,\text{kpc}\). Then we cut these in half again, into rings of \(4\,\text{kpc}\), then into eight rings of \(2\,\text{kpc}\) and finally, into 1\,kpc rings. We note that cutting the model into smaller pieces leads to a better radial resolution but at the same time increased uncertainties and correlations between components (see Fig.\,\ref{fluxes}). Therefore, we choose a 2\,kpc binning until 10\,kpc, and one 5\,kpc bin from 10--15\,kpc, given that the uncertainties on the fluxes become orders of magnitude larger than inside the Solar circle. The GCE model calculations are only available up to 15\,kpc. {In the Cases 1--4}, for all cut models the best fit value combination is found to be \(R_E=5.0\,\)kpc and \(h_E=0.4\,\)kpc, corresponding to  $V_\text{eff}=$126\,kpc$^3$ for the entire Galaxy and 101\,kpc$^3$ up to 15\,kpc. {The likelihood of a cut Spiral Arm model (Case 5) is much better compared to the global model but still performs worse than Cases 1--4.}
Beyond the \(10\,\text{kpc}\) ring, all fluxes are consistent with zero within uncertainties. This also makes sense considering that the stellar population in the outermost part of the thin disk is old and shows low metallicity, conditions that we generally do not associate with \nuc{Al}{26}. We use the last bin to fully constrain the mass of the \nuc{Al}{26} in the Galaxy, but which has no impact on the comparison between massive stars and novae.

\begin{table*}[h!]
\caption{Model compositions used in a $\chi^2$ test in order to determine the relative contribution of $M_\text{MS}$ and $M_\text{N}$  to Galactic \nuc{Al}{26}.}                 
\label{contribution_Modelle}    
\centering                        
\begin{tabular}{c | l }      
\hline\hline               
 Model & Model $M_\text{N}$ + Model $M_\text{MS}$ \\         
\hline                      
    VV  & CN \citep{vasini2024tracingmilkywayspiral} + massive stars \citep{vasini2024tracingmilkywayspiral}\\
  VMa & CN \citep{vasini2024tracingmilkywayspiral} + non-rot VMS<300\,M$_\odot$\citep{Martinet_2022} \\
    VMb & CN \citep{vasini2024tracingmilkywayspiral} + rot VMS<300\,M$_\odot$\citep{Martinet_2022}\\
     VMc & CN \citep{vasini2024tracingmilkywayspiral} + non-rot VMS<500\,M$_\odot$\citep{Martinet_2022}\\
       VMd & CN \citep{vasini2024tracingmilkywayspiral} + rot VMS<500\,M$_\odot$\citep{Martinet_2022}\\
       \hline
\end{tabular}
\end{table*}

Fig.\,\ref{MassModels} shows the conversion of the measured 1.8\,MeV line fluxes into persistent \nuc{Al}{26} masses for the best fit combination, \(R_E=5.0\,\) kpc and \(h_E=0.4\,\)kpc, of the different models. Within statistical uncertainties, the masses per ring are consistent with each other. The spread of the absolute values then provides an estimate of the systematic uncertainties. When compared to the total model from \citet{vasini2024tracingmilkywayspiral}, they show generally good agreement without adjusting for any relative amplitudes. 

Comparison with \citet{Martinet_2022} shows that the non-rotating model for VMS up to 300\,M${_\odot}$ is not consistent with the observed abundance of \nuc{Al}{26} throughout all of the Galactic rings. In the case of rotating VMS with the same upper mass boundary, we see barely consistent masses up to 6\,kpc and misalignment for the outer regions of the Galaxy. For higher VMS, the agreement between observed and predicted abundances becomes generally good for both rotating and non-rotating models, within 8\,kpc distance from the Galactic center. 

We find that in the 8--10\,kpc bin, no model expectations align with the measurements, as could be expected from previous findings. The local environment around the Sun shows somewhat enhanced 1.8\,MeV line flux due to the numerous nearby young massive stars regions \citep[e.g.,][]{Diehl2010_ScoCen,Martin2012_posi,Siegert2017_MassiveStarRegions, Krause2018_SuperbubbleScoCen} and potentially from the Local Bubble \citep[e.g.,][]{schulreich23_localBubble, Siegert2024_localBubble}.
Because star formation is a stochastic process, this local enhancement, while related to the SFR, may appear by chance \citep{Combinations, dissertation_Pleintinger, Siegert2023_GalacticPopulation}.
With a sensitive enough Compton telescope, such as COSI \citep{Tomsick2019_COSI}, such a foreground could be identified, so that a more-precise model prediction can be tested.

\section{Relative contribution of classical novae and massive stars}\label{contribution_results}
If we consider the total model \nuc{Al}{26} mass $M$ to be a linear combination, with parameters $a_\text{N}$ and $a_\text{MS}$, of the masses originating in massive stars and classical novae, $M_\text{MS}$ and $M_\text{N}$ respectively,
\begin{align}
    M(a_\text{N},a_\text{MS})= a_\text{N}\times M_\text{N} + a_\text{MS}\times M_{\rm{MS}}
    \label{Chifit}
\end{align}
we can fit it for the mass distribution predicted by a geometrical model via a $\chi^2$-test, because the measurement uncertainties $\sigma_{26}(r_{ij})$ are approximately Gaussian,

\begin{align}
    \chi^2=\sum_{i,j=1}^{N} \left( \frac{ M_{26}(r_{ij})-M(r_{ij},a_\text{N},a_\text{MS})}{\sigma_ {26}(r_{ij})}\right)^2\mathrm{.}
    \label{eq:chi2}
\end{align}

\begin{figure}[!ht]
    \centering
    \includegraphics[width=\linewidth]{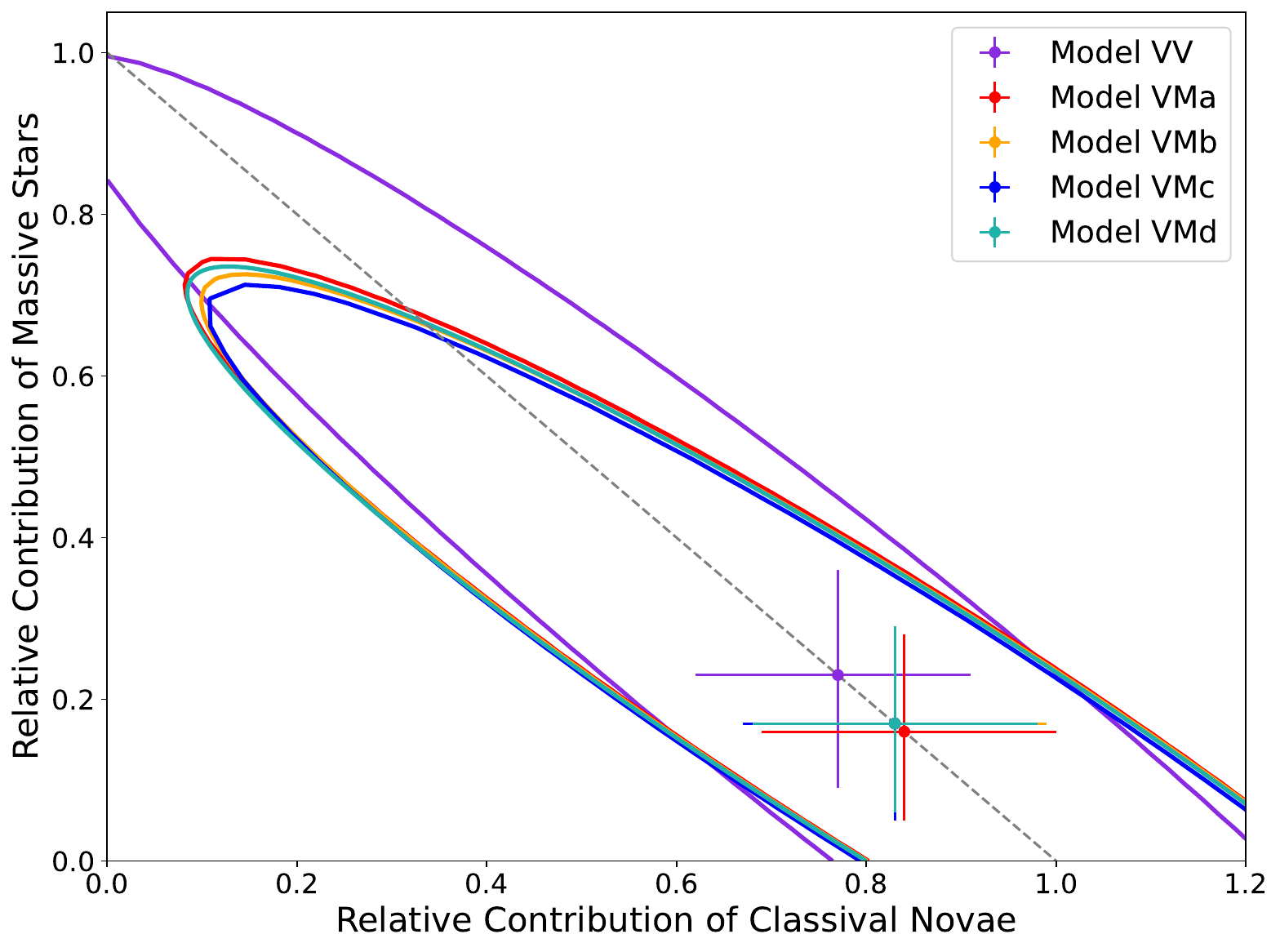}
    \caption{Results of a $\chi^2$ test for the best spatial models to estimate their confidence intervals. For comparison the 1$\sigma$ values (crosses) are shown with respect to the $2\sigma$ ellipses for selected model combinations. The extent of the full two-parameter confidence regions is larger than the marginalised intervals, because of the anticorrelation between the two source amplitudes. Due to the nature of the measurement, also the absolute \nuc{Al}{26} carries an uncertainty, so that the relative contributions per source type can be more than $1$.}
    \label{Results_plot}
\end{figure}

Eq.\,(\ref{eq:chi2}) is optimized for $a_\text{N}$ and $a_\text{MS}$ of the five different model combinations as listed in Tab.\,\ref{contribution_Modelle}. We use the data of the doubly-exponential disk from Fig.\,\ref{MassModels} for illustration and note that the results of the other disk models are similar.
After {absolute} normalization to the best fit model, the parameters $a_\text{N}$ and $a_\text{MS}$ give the relative combination of source contributions. Their uncertainties are determined by confidence intervals obeying $\Delta \chi^2 = 2.3$ for $1\sigma$ uncertainties and {$6.2$ for $2\sigma$}, respectively.
We distinguish between the joint two-dimensional confidence regions in $(a_{\rm N},a_{\rm MS})$, shown in Fig.\,\ref{Results_plot}, and the one-dimensional marginalised distributions of the individual source fractions, summarized in Tab.\,\ref{results_Contribution}.
Because the two source amplitudes are anticorrelated, the projection of the joint confidence region onto either parameter axis is substantially broader than the corresponding confidence interval obtained from its marginalised one-dimensional distribution.
We test five different models, see Table\,\ref{contribution_Modelle}, considering the GCE by \citet{vasini2024tracingmilkywayspiral} as is (Model\,VV), but also combining it with the model of \cite{Martinet_2022} (Models\,VMa--d) in order to account for possible contributions of VMS to Galactic \nuc{Al}{26}, that could diminish any novae contributions.

We note that the geometrical distribution of VMS in \citet{Martinet_2022} are very similar, so that Models\,VMa--d can be considered only as slight variations.
We find good agreement between all tested models and show the results of the best fit spatial models in Fig.\,\ref{Results_plot}. 
All tested models predict a non-negligible contribution of novae in the $1\sigma$ interval. The ellipse showing the confidence interval of Model\,VV still includes contributions of up to 70\,\% massive stars in its outmost region, while all other models predict considerably less.

Alternatively, we determine the amplitudes in a constrained fit where the total mass is a pure combination of Model\,1 and Model\,2 with $1 = a_\mathrm{N} + a_\mathrm{MS}$, $0 \leq a_1 \leq 1$, and $0 \leq a_2 \leq 1$.
This results in amplitudes in the range $a_\mathrm{N} \in [0.31,0.85]$ and $a_\mathrm{MS} \in [0.15,0.69]$.
Finally, we perform a maximum likelihood ratio test where we set $a_\mathrm{N}$ to zero in order to estimate how significant the nova contribution is.
Typical values for $2\Delta\ln\mathcal{L}$ are $5$ or more, which suggest a corrected $p$-value of $0.013$, taking into account that the null-hypothesis sets the nova contribution to the boundary of the physically-allowed range.
This is equivalent to a ``significance of a nova contribution'' of at least $2.5\sigma$.
We note, however, that the maximum likelihood fit for the case of massive stars alone is not adequate.

In general, the confidence interval for Model\,VV shows the steepest slope. That means, it includes more combinations in favour of higher massive star contribution than the other models, although marginally.
Models\,VMa--d are almost identical, as explained above, even though the total mass per ring can vary substantially. 
This is not surprising as $\chi^2$ only depends on the relative mass distribution, which is very similar for all \citet{Martinet_2022} predictions.

For the joint two-dimensional $2\sigma$ confidence regions shown in Fig.\,\ref{Results_plot}, Model VV extends to zero nova contribution at the outer edge of the contour, while the VMa--d models extend to approximately 8\%.
These values should not be confused with the $2\sigma$ lower bounds of the marginalised one-dimensional nova distributions, which are 52--63\% for the baseline analysis (Tab.\,\ref{results_Contribution}).
These values exceed previous contribution estimates by far and are not in agreement with the existing literature \citep[e.g.,][]{Bennett_2013, jose1998nuclearuncertaintiesnenamgalcycles}.

{We additionally want to test how the enhanced local foreground might influence these results. For a purely geometrical approach we exclude the bin between 8--10\,kpc from the $\chi^2$-test. This shifts the Models VMa--d to lower nova contributions, while Model VV is shifted to even higher contributions. The minimal nova contribution then is between (0--2)\% within 2$\sigma$ and (7--14)\% within 1$\sigma$ for Models VMa--d and 65\% and 75\% within $2\sigma$ and $1\sigma$, respectively. This is because excluding the bin changes the steepness of the distribution beyond the peak depending on the source types. In the case of VMa--d, this favours a later peak of the best fit distribution and therefore a higher VMS contribution. In case of VV, the steeper distribution requires a higher contribution of classical novae. A second, less drastic test is to increase the uncertainty within this bin, in order to account for possible systematic uncertainties. This also reduces the minimum mass contribution of classical novae depending on the scaling factor. For factors (1.5, 2, 2.5, 3, 5) we find at least (0.13, 0.08, 0.03, 0, 0) nova contribution within 2$\sigma$ for VMa--d and (0.64, 0.65, 0.65, 0.65, 0.65) for VV within the same level of uncertainty. By scaling up the uncertainties, the weight of the bin is shifted in the $\chi^2$-test and the 8--10\,kpc becomes less influential. High enough scaling factors ($\approx5)$ are therefore equivalent to leaving the bin out completely. This is therefore a softer approach than just leaving the bin out.}

{A more physical approach is to reduce the flux in the 8--10\,kpc bin by an estimate for a local foreground component from the Local Bubble of $\sim 10^{-5}$\,ph\,s$^{-1}$\,cm$^{-2}$ as estimated by \citet{Siegert2024_localBubble}. This leaves the results for Model VV unchanged but decreases nova contributions for Models VMa--d to at least 21--27\% within 2$\sigma$, or 31--38\% within 1$\sigma$.
{These tests show that, although the preferred solutions remain nova rich, a lower bound of $\gtrsim 50\%$ is not robust to the treatment of the 8--10\,kpc bin for the VMa–d source profiles.
Model VV is considerably less sensitive to these tests and remains nova dominated.}
This is somewhat more in line with the results of \citet{Bennett_2013} within uncertainties but still exceed the estimates of \citet{jose1998nuclearuncertaintiesnenamgalcycles}.}

\begin{table*}[]
\caption{Best-fit relative source contributions}
    \centering
    \begin{tabular}{c|c c c c c | c c c c c}
\hline\hline
 & \multicolumn{5}{c}{Classical Novae} & \multicolumn{5}{c}{Massive Stars} \\
Model &-2$\sigma$  &$-\sigma$& Best & $+\sigma$ &+2$\sigma$&-2$\sigma$&  $-\sigma$ & Best & $+\sigma$ & +2$\sigma$\\
\hline
Model\,VV & 0.52 &0.62 & 0.77 & 0.91 & 1.0 & 0.00$^{(b)}$ & 0.13 & 0.23 & 0.37 & 0.45 \\
Model\,VMa & 0.63 & 0.69& 0.84 & 1.00 & 1.0$ ^{(a)}$ & 0.00$^{(b)}$ & 0.05 & 0.16 & 0.28 & 0.36 \\
Model\,VMb & 0.55 & 0.67& 0.83 & 0.99 & 1.0 $ ^{(a)}$& 0.00$^{(b)} $& 0.05 & 0.17 & 0.29 & 0.36 \\
Model\,VMc & 0.54 & 0.67& 0.83 & 0.98 & 1.0 $ ^{(a)}$& 0.00$^{(b)}$ &0.06 & 0.17 & 0.28 & 0.38\\
Model\,VMd & 0.54 & 0.67& 0.83 & 0.98 & 1.0 $^{(a)}$ & 0.01 &0.06 & 0.17 & 0.29 & 0.33\\
\hline
\end{tabular}
   
    \tablefoot{Best-fit relative source amplitudes and marginalised $1\sigma$ and $2\sigma$ intervals through the $\chi^2$ minimum for a doubly-exponential disk model. These intervals are not the extrema of the parameters over the full joint confidence regions. Results of other data sets look similar. (a) Values clipped at 1.0. (b) Values clipped at 0.0.}
    \label{results_Contribution}
\end{table*}

\section{Summary and conclusion}\label{summary}
In this paper, we determine the Galactic mass distribution of \nuc{Al}{26}, and, by assuming a linear combination of source populations, the contribution of classical novae and massive stars to this budget. 
Using line-of-sight integration, maps of the \nuc{Al}{26} emission for different spatial models are created and fitted to a 20\,yr INTEGRAL/SPI dataset. By fitting six maps, representing concentric rings throughout the Galaxy, to the data, the mass distribution depending on the Galactocentric distance is determined. A doubly-exponential disk model with values $R_E=5.0\,$kpc and $h_E=0.4\,$kpc was determined to fit best, resulting in a total \nuc{Al}{26} mass of $3.3$--$4.6\,\mathrm{M_\odot}$ within 15\,kpc. 
\citet{vasini2024tracingmilkywayspiral} present a theoretical Galactic distribution of \nuc{Al}{26}, depending on the contributions of massive stars and classical novae. We find good agreement between our measurements of the mass distribution for all spatial models with their predictions. Only the measurement bin between 8--10\,kpc shows a somewhat enhanced mass, which may be attributed to the numerous amount of young stars in this region as well as potential contributions of the Local Bubble \citep[e.g.,][]{schulreich23_localBubble, Siegert2024_localBubble}. By assuming that the total mass is a linear combination of both source types, we find $77^{+14}_{-15}\,\%$ nova and $23^{+13}_{-14}\%$ massive star contributions when our data is compared to the GCE by \citet{vasini2024tracingmilkywayspiral}.
.
{This baseline result is substantially larger than previous estimates of about 10\% nova contribution \citep[e.g.,][]{Jose_98_ONeOCnovae}.}
{Considering a local foreground component from the Local Bubble alleviates this somewhat within uncertainty but not enough to explain the Galactic distribution of \nuc{Al}{26} without a significant contribution of classical novae.}
As \nuc{Al}{26} might be ejected from VMS, they are considered as a possible source as well. We compare the mass distributed by novae \citep{vasini2024tracingmilkywayspiral} and the distribution of \nuc{Al}{26} by VMS \citep{Martinet_2022}.
{The baseline fits using the VMS profiles likewise favour large nova fractions.
However, their lower confidence limits are particularly sensitive to the treatment of the 8--10\,kpc bin:
Excluding this bin permits nova fractions approaching zero at $2\sigma$, whereas subtracting an estimated Local Bubble foreground gives lower limits of 21--27\%.}

In conclusion, the contribution of classical novae \nuc{Al}{26} may be much larger than previously estimated.

Under our baseline assumptions, the best-fit source decompositions favour a dominant nova contribution.
However, the full confidence regions are broad, and for some source profiles and/or assumptions on the 8--10\,kpc bin, a massive-star-dominated or even massive-star-only solution remains allowed at the $2\sigma$ level.
If the nova models are true, however, this alleviates the tension in SFR measurements from \nuc{Al}{26} drastically. {Taking the baseline VV best-fit fraction of 23\% from massive stars at face value would imply a SFR} of $1.2$--$1.8\,\mathrm{M_\odot\,yr^{-1}}$. {Under the same assumption,}, it increases the \nuc{Fe}{60}/\nuc{Al}{26} ratio from massive stars and ccSNe to $0.8\pm0.2$, as less \nuc{Al}{26} would origin in ccSNe. This is more in line with theoretical predictions \citep[e.g.,][]{Spyrou_nature, Timmes_1995}.

While current MeV $\gamma$-ray observations cannot identify individual massive stars nor classical novae at 1.8\,MeV, future observations, such as with COSI \citep[Compton Spectrometer and Imager,][]{Tomsick2019_COSI}, which is planned for launch in 2027, may for the first time identify massive stars.
The best candidates would be nearby pre-supernova stars with high mass loss rates and massive star binaries, such as $\gamma^2$-Velorum \citep[zero-age main sequence mass $\sim 30\,\mathrm{M_\odot}$, distance $340$\,pc][]{Oberlack2000_gamvel,dissertation_Pleintinger}, Rigel ($\sim 25\,\mathrm{M_\odot}$, $260$\,pc), Betelgeuse ($\sim 20\,\mathrm{M_\odot}$, $200$\,pc), or S\,Monocerotis\,A\&B ($20$--$30\,\mathrm{M_\odot}$, $300$--$700$\,pc).
Limits on the 1.809\,MeV flux from some of these objects have recently been evaluated by \citet{Mittal2026_ALPs}.
The typical scale of the mass limits is on the order of $10^{-4}\,\mathrm{M_\odot}$, whereas typical wind yields are on the order of $10^{-7}$--$10^{-5}\,\mathrm{M_\odot}$ for initial stellar masses up to $30\,\mathrm{M_\odot}$ \citep{dissertation_Pleintinger}.

\begin{acknowledgements}
      Saurabh Mittal acknowledges support by the  Bundesministerium für Wirtschaft und Energie via the Deutsches Zentrum für Luft- und Raumfahrt (DLR) under Contract No. {50\,OO\,2219}. Laura Eisenberger acknowledges support by the Bundesministerium für Wirtschaft und Energie via the Deutsches Zentrum für Luft- und Raumfahrt (DLR) under Contract No. {50\,OR\,2413} and is grateful for the support of the {Stu\-di\-en\-stif\-tung des Deu\-tschen Vol\-kes}. Dimitris Tsatsis and Tristan Bouchet acknowledge support from the {DFG/LIS} project {SI\,2502/6-1, project number 551127478}. Rudi Reinhardt acknowledges support by the \emph{Bun\-des\-mi\-nis\-te\-ri\-um für Forschung, Technologie und Raumfahrt} as part of the \emph{FORKA (Forschung für den Rückbau kerntechnischer Anlagen)} funding program under the reference number {15S9455\,A-E}. Tomohiko Oka acknowledges support by the JSPS Overseas Research Fellowships. Manuel Skalka acknowledges support by the  {Bundesministerium für Wirtschaft und Energie via the Deutsches Zentrum für Luft- und Raumfahrt (DLR)} under Contract No. {50\,OR\,2522}.
      Arianna Vasini acknowledges financial support from the European Research Council under the {ERC Starting Grant} ``GalFlow'' (grant {101116226}). We thank S\'ebastien Martinet for providing their very massive star models.
\end{acknowledgements}

\bibliographystyle{aa}
\bibliography{refs}

\end{document}